%% file: CIKM2020-SMITH-Text-Match.tex
\documentclass[sigconf, balance=false]{acmart}
\usepackage{booktabs} 

\setcopyright{rightsretained}
\usepackage[english]{babel}
\usepackage{subcaption}
\usepackage[inline]{enumitem}
\usepackage{amsmath}
\usepackage{stmaryrd}
\usepackage{graphicx}
\usepackage{subcaption}
\usepackage{multirow}
\usepackage[utf8]{inputenc}
\usepackage{color,soul}
\usepackage{dsfont}

\usepackage{balance}
\AtBeginEnvironment{thebibliography}{\balance}

\def\ignore#1{}

\newcommand{\astfootnote}[1]{
\let\oldthefootnote=\thefootnote
\setcounter{footnote}{0}
\renewcommand{\thefootnote}{\fnsymbol{footnote}}
\footnote{#1}
\let\thefootnote=\oldthefootnote
}

\begin{document}
\title{Beyond 512 Tokens: Siamese Multi-depth Transformer-based Hierarchical Encoder for Long-Form Document Matching}


\author{Liu Yang \quad Mingyang Zhang \quad Cheng Li \quad Michael Bendersky \quad Marc Najork}

\affiliation{%
	\institution{
		Google Research, Mountain View, CA, USA  \\}
}
\email{{yangliuy,mingyang,chgli,bemike,najork}@google.com}

\begin{abstract}
	\noindent Many natural language processing and information retrieval problems can be formalized as the task of semantic matching. Existing work in this area has been largely focused on matching between short texts (e.g., question answering), or between a short and a long text (e.g., ad-hoc retrieval). Semantic matching between long-form documents, which has many important applications like news recommendation, related article recommendation and document clustering, is relatively less explored and needs more research effort. In recent years, self-attention based models like Transformers \cite{NIPS2017_Transformers} and BERT \cite{devlin2018bert} have achieved state-of-the-art performance in the task of text matching. These models, however, are still limited to short text like a few sentences or one paragraph due to the quadratic computational complexity of self-attention with respect to input text length. In this paper, we address the issue by proposing the Siamese Multi-depth Transformer-based Hierarchical (SMITH) Encoder for long-form document matching. Our model contains several innovations to adapt self-attention models for longer text input. We propose a transformer based hierarchical encoder to capture the document structure information.
	In order to better capture sentence level semantic relations within a document, we pre-train the model with a novel masked sentence block language modeling task in addition to the masked word language modeling task used by BERT. Our experimental results on several benchmark datasets for long-form document matching show that our proposed SMITH model outperforms the previous state-of-the-art models including hierarchical attention~\cite{Yang:2016:ARS:2983323.2983818}, multi-depth attention-based hierarchical recurrent neural network~\cite{10.1145/3308558.3313707}, and BERT. Comparing to BERT based baselines, our model is able to increase maximum input text length from 512 to 2048. We will open source a Wikipedia based benchmark dataset, code and a pre-trained checkpoint to accelerate future research on long-form document matching.\footnote{The code and a pre-trained checkpoint of the proposed SMITH model will be available at \url{https://github.com/google-research/google-research/tree/master/smith}. The Wikipedia based benchmark dataset will be available at \url{https://github.com/google-research/google-research/tree/master/gwikimatch}. Please note that different from the Wikipedia dataset used in this paper, the released dataset will contain both machine generated document pairs and human annotated document pairs. We hope the dataset can be a useful open benchmark for future research on document matching.}
\end{abstract}



%





\copyrightyear{2020}
\acmYear{2020}
\acmConference[CIKM '20]{Proceedings of the 29th ACM International Conference on Information and Knowledge Management}{October 19--23, 2020}{Virtual Event, Ireland}
\acmBooktitle{Proceedings of the 29th ACM International Conference on Information and Knowledge Management (CIKM '20), October 19--23, 2020, Virtual Event, Ireland}\acmDOI{10.1145/3340531.3411908}
\acmISBN{978-1-4503-6859-9/20/10}

\fancyhead{}
\settopmatter{printacmref=true, printfolios=false}

\maketitle

\input{intro}
\input{related}
\input{method}

\input{exp_part1}
\input{exp_part2}
\input{conclusion}

\bibliographystyle{ACM-Reference-Format}
\bibliography{reference}

\end{document}

%% file: intro.tex
\section{Introduction}
\label{sec:intro}



Semantic matching is an essential task for many natural language processing (NLP) and information retrieval (IR) problems. Research on semantic matching can potentially benefit a large family of applications including ad-hoc retrieval, question answering and recommender systems \cite{10.5555/2683840}. Semantic matching problems can be classified into four different categories according to text length, including short-to-short matching, short-to-long matching, long-to-short matching and long-to-long matching. Table \ref{tab:different_text_match_task} shows a classification of different semantic matching tasks with example datasets. Semantic matching between short text pairs is relatively well studied in previous research on paraphrase identification \cite{DBLP:conf/naacl/YinS15}, natural language inference \cite{bowman-etal-2015-large}, answer sentence selection \cite{yang-etal-2015-wikiqa}, etc. Short-to-long semantic matching like relevance modeling between query/ document pairs has also been a popular research topic in IR and NLP communities \cite{craswell2020overview}. For long-to-short semantic matching, there are also a variety of research on tasks like conversation response ranking, which is to match a conversation context with response candidates \cite{DBLP:journals/corr/LowePSP15}. To the best of our knowledge, semantic matching between long document pairs, which has many important applications like news recommendation, related article recommendation and document clustering, is less explored and needs more research effort. Table \ref{tab:doc_match_example} shows an example of semantic matching between document pairs from  Wikipedia. These documents have thousands of words organized in sections, passages and sentences.

\begin{table*}[]
	\small
	\caption{A classification of different semantic matching tasks. The focus of this paper is \textit{long-to-long} document matching. }
	\vspace{-0.15in}
	\label{tab:different_text_match_task}
\begin{tabular}{p{1.5cm}|p{3cm}|p{2cm}|p{9cm}}
\hline \hline
Type                                                   & Tasks                           & Example Data                       & Explanation                                                                                                  \\ \hline
\multirow{3}{*}{Short-to-short}               & Paraphrase Identification       & MRPC \cite{dolan-brockett-2005-automatically}                          & Given two sentences, predict whether they have the same semantic meaning.                                   \\ \cline{2-4} 
                                                       & Answer Sentence Selection & WikiQA \cite{yang-etal-2015-wikiqa}         & Given a question and candidate answer sentences, select a correct answer. \\ \cline{2-4} 
                                                       & Textual Entailment  & SNLI \cite{bowman-etal-2015-large}                         & Given two sentences, predict whether they have textual entailment relations.             \\ \hline
\multirow{2}{*}{Short-to-long}                & Document Ranking                & TREC 2019 Deep Learning \cite{craswell2020overview} & Given a query and a candidate document set, rank documents according to query/ document relevance.        \\ \cline{2-4} 
                                                       & Blog Search                    & TREC 2008 Blog Track \cite{DBLP:conf/trec/OunisMS08}
                                                                    & Given a query and a blog collection, rank blog posts according to topic relevance or opinions. \\ \hline
\multirow{1}{*}{Long-to-short}                & Response Ranking   & UDC \cite{DBLP:journals/corr/LowePSP15}           & Given a dialog context and candidate responses, select a high quality response.               \\\hline
\multirow{2}{*}{\textbf{\textit{Long-to-long}}} & Related Article Suggestion & Wikipedia \cite{10.1145/3308558.3313707}                     & Given a document pair, predict whether they are relevant with each other.                                   \\ \cline{2-4} 
                                                       & Paper Citation Suggestion       & AAN \cite{10.5555/1699750.1699759}                           & Given a paper pair, predict whether one paper is a good citation of the other.               \\ \hline \hline
\end{tabular}
\end{table*}

\begin{table*}[t]
	\small
	\caption{An example to illustrate the document matching task from the Wikipedia data. ``Sim'' means the similarity estimated based on the Jaccard similarity between the outgoing links of two documents. ``Len'' means the document length by words.}
	\vspace{-0.15in}
	\label{tab:doc_match_example}
\begin{tabular}{l|p{3.5cm}|p{1.8cm}|p{7cm}|l|l|l}
\hline \hline
Type       & URL                                                    & Title                   & Document Content & Label            & Sim       & Len \\ \hline
Source & \url{http://en.wikipedia.org/wiki/Chartered\_Engineer\_(UK)} & Chartered Engineer (UK) & In the United Kingdom, a Chartered Engineer is an Engineer registered with the Engineering Council (the British ...... & \textbackslash{} & \textbackslash{} & 1147            \\ \hline
Target & \url{http://en.wikipedia.org/wiki/Engineering\_Council}      & Engineering Council     & The Engineering Council (formerly Engineering Council UK; colloquially known as EngC) is the UK's regulatory ...... & 1    & 0.5846           & 999             \\ \hline
Target & \url{http://en.wikipedia.org/wiki/Institute\_of\_Physics}    & Institute of Physics    & The Institute of Physics (IOP) is a UK-based learned society and professional body that works to advance physics ...... & 0 & 0.0099           &       2036          \\ \hline \hline
\end{tabular}
\end{table*}

Compared to semantic matching between short texts, or between short and long texts, semantic matching between long texts is a more challenging task due to a few reasons: 1) When both texts are long, matching  them requires a more thorough understanding of semantic relations including matching pattern between text fragments with long distance; 2) Long documents contain internal structure like sections, passages and sentences. For human readers, document structure usually plays a key role for content understanding. Similarly, a model also needs to take document structure information into account for better document matching performance; 3) The processing of long texts is more likely to trigger practical issues like out of TPU/GPU memories without careful model design. In the recent two years, self-attention based models like Transformers \cite{NIPS2017_Transformers} and BERT \cite{devlin2018bert} have achieved the state-of-the-art performance in several natural language understanding tasks like sentence pair classification, single sentence classification and answer span detection. These kinds of models, however, are still limited to the representation and matching of short text sequences like sentences due to the quadratic computational time and space complexity of self-attention with respect to the input sequence length \cite{NIPS2017_Transformers}. To handle this challenge, we would like to design a long-form text encoder, which combines the advantages of sequence dependency modeling with self-attention in Transformers and long text processing with hierarchical structures for document representation learning and matching. 



Perhaps the closest prior research to our work is the study on the semantic text matching for long-form documents by Jiang et. al. \cite{10.1145/3308558.3313707}. They proposed the MASH RNN model to learn document representations from multiple abstraction levels of the document structure including passages, sentences and words. However, the adopted attentive RNN component in the MASH RNN model may suffer from the gradient vanishing and explosion problems on long input sequences. It is difficult for RNN based models to capture the long distance dependencies in long documents, which might lead to sub-optimal performance on long text content modeling compared with self-attention models like Transformers/ BERT where there are direct interactions between every token pair in a sequence.
It is the right time to revisit this line of work and further push the boundary of long text content understanding with self-attention models like Transformers. However, as presented before, building Transformer based long text encoder is not trivial because of the quadratic computational time and memory complexity of self-attention with respect to the input sequence length. For example, the maximum input text length of BERT is 512 for single sentence classification, and less than 512 for sentence pair classification.



We address these issues by proposing the Siamese Multi-depth Transformer-based Hierarchical (SMITH) Encoder for document representation learning and matching, which contains several novel design choices to adapt self-attention models like Transformers/ BERT for modeling long text inputs. Our proposed text matching model adopts a two-tower structure of Siamese network, where each tower is a multi-depth Transformer-based hierarchical encoder to learn the document representations. We first split the input document into several sentence blocks, which may contain one or more sentences using our proposed greedy sentence filling method. Then the sentence level Transformers learn the contextual representations for the input tokens in each sentence block. We represent the whole sentence block with the contextual representations of the first token, following the practice in BERT. Given a sequence of sentence block representation, the document level Transformers learn the contextual representation for each sentence block and the final document representation. This model design brings several benefits in terms of model training and serving: 1) The Siamese model architecture is a better choice to serve with efficient similarity search libraries for dense vectors\cite{JDH17,NIPS2017_7157}, since document representations can be generated independently and indexed offline before online serving. 2) The hierarchical model can capture the document internal structural information like sentence boundaries. 3) Compared with directly applying Transformers to the whole document, the two level hierarchical SMITH model including sentence level and document level Transformers reduces the quadratic memory and time complexity by changing the full self-attention on the whole document to several local self-attentions within each sentence block. The sentence level Transformers capture the interactions between all token pairs within a sentence block, and the document level Transformers maintain the global interaction between different sentence blocks for long distance dependencies. 

Inspired by the recent success of language model pre-training methods like BERT, SMITH also adopts the ``unsupervised pre-training + fine-tuning'' paradigm for the model training. For the model pre-training, we propose the masked sentence block language modeling task in addition to the original masked word language modeling task used in BERT for long text inputs. When the input text becomes long, both relations between words in a sentence block and relations between sentence blocks within a document becomes important for content understanding. Therefore, we mask both randomly selected words and sentence blocks during model pre-training. The sum of the masked sentence block prediction loss and the masked word prediction loss is the final SMITH model pre-training loss. The model fine-tuning process is similar to BERT, where we remove the word/ sentence level masks and fine-tune the model parameters initialized with pre-trained checkpoints with only the text matching loss. We evaluate the proposed model with several benchmark data for long-form text matching \cite{10.1145/3308558.3313707}. The experimental results show that our proposed SMITH model outperforms the previous state-of-the-art Siamese matching models including hierarchical attention~\cite{Yang:2016:ARS:2983323.2983818}, multi-depth attention-based hierarchical recurrent neural network~\cite{10.1145/3308558.3313707}, and BERT for long-form document matching, and increases the maximum input text length from 512 to 2048 when compared with BERT-based baselines.



 






Our main contributions can be summarized as follows:

 \begin{itemize}
	\item We propose the Siamese Multi-depth Transformer-based Hierarchical (SMITH) Encoder for document matching, which contains several novel design choices to adapt self-attention models for modeling long text inputs. 
	\item For model pre-training, we propose  the masked sentence block language modeling task to capture sentence level semantic relations within a document towards better long text content understanding.
	\item Experimental results on several benchmark data for long-form text matching \cite{10.1145/3308558.3313707} show that our proposed SMITH model outperforms the previous state-of-the-art models and increases the maximum input text length from 512 to 2048 when comparing with BERT based baselines. We will open source a Wikipedia based benchmark dataset, code and a pre-trained model checkpoint to accelerate future research on document understanding and matching.
\end{itemize}





%% file: related.tex
\section{Related Work}
\label{sec:rel}


\textbf{Neural Matching Models.}
A number of neural matching models have been proposed for information retrieval and natural language processing \cite{DBLP:conf/cikm/HuangHGDAH13,DBLP:conf/nips/HuLLC14,DBLP:conf/aaai/PangLGXWC16,Guo:2016:DRM:2983323.2983769,Yang:2016:ARS:2983323.2983818,DBLP:conf/acl/WuWXZL17, Xiong:2017:ENA:3077136.3080809,Mitra:2017:LMU:3038912.3052579,alime-tl}. These models can be classified into the \textit{representation-focused} models and the \textit{interaction-focused} models~\cite{Guo:2016:DRM:2983323.2983769,DBLP:journals/corr/abs-1903-06902}. The \textit{representation-focused} models learn the representations of queries and documents separately, and then they measure the similarity of the representations with functions like cosine, dot, bilinear or tensor layers. On the other hand, the \textit{interaction-focused} models build a query-document word pairwise interaction matrix to capture the exact matching and semantic matching information between query-document pairs. Then a deep neural network which can be a CNN \cite{DBLP:conf/nips/HuLLC14,DBLP:conf/aaai/PangLGXWC16,alime-tl}, term gating network with histogram or value shared weighting mechanism \cite{Guo:2016:DRM:2983323.2983769,Yang:2016:ARS:2983323.2983818} is applied to the query-document interaction matrix to generate the final ranking score. There are also neural matching models which combine the ideas of the \textit{representation-focused} and \textit{interaction-focused} models \cite{Mitra:2017:LMU:3038912.3052579,alime-tl}. 
Since it is difficult to serve \textit{interaction-focused} models for online fast inference due to enormous computational costs of the interaction matrices for all query-document pairs, our proposed model belongs to \textit{representation-focused} models, which are also called Siamese models or ``Dual Encoder'' models.


\textbf{Self-Attention Models for Long Text Modeling.}
Self-attention models like Transformer and BERT show promising performance on several tasks in natural language processing and information retrieval. Most of these models are restricted to the representation and matching of short text sequences like sentences and passages. Our work is also built on top of Transformers with a different focus on effective representation learning and matching of long text. Recently there are some related works on adapting Transformers for long text modeling \cite{kitaev2020reformer,qiu2019blockwise,DBLP:journals/corr/abs-1905-06566,DBLP:journals/corr/abs-1901-02860,DBLP:journals/corr/abs-1906-08237,DBLP:journals/corr/abs-1905-07799,roy2020efficient,DBLP:journals/corr/abs-1904-10509,rae2019compressive,ho2019axial}. Zhang et al. \cite{DBLP:journals/corr/abs-1905-06566} proposed the HiBERT model for document summarization and a method to pre-train it using unlabeled data. Our work on the SMITH model is inspired by their research with several differences. First we split the document into sentence blocks with greedy sentence filling for more compact input text structures and less padded words. Second we build a Siamese ``Dual Encoder'' model for document pair similarity modeling. Third we propose a novel pre-training task based on dynamic masked sentence block prediction instead of predicting the masked sentence one word per step as in \cite{DBLP:journals/corr/abs-1905-06566}. We also consider combining representations from different levels of hierarchical Transformers for richer representations.


\textbf{Unsupervised Language Model Pre-training.} The idea of unsupervised learning from plain text for language model pre-training has been explored in several works like Word2Vec\cite{DBLP:conf/nips/MikolovSCCD13}, ELMo\cite{DBLP:journals/corr/abs-1802-05365}, GPT\cite{Radford2018ImprovingLU} and BERT\cite{devlin2018bert}. These models can be pre-trained by predicting a word or a text span using other words within the same sentence. For example,  Word2Vec can be trained by predicting one word with its surrounding words in a fixed text window and BERT pre-trains a language model by predicting masked missing words in a sentence given all the other words. We also study model pre-training techniques on plain text to improve the downstream document matching task. In addition to the masked word prediction task in BERT, we propose the masked sentence block prediction task to learn the relations between different sentence blocks. 

%% file: method.tex
\section{Method Overview}
\label{sec:method_overview}

\subsection{Problem Formulation}


We define the task of document matching following previous literature \cite{10.1145/3308558.3313707}. We are given a source document $d_s$ and a set of candidate documents $\mathcal{D}_c$. The system needs to estimate the semantic similarities $\hat{y} = Sim(d_s, d_c)$, where $d_c \in \mathcal{D}_c$, for every document pair $(d_s, d_c)$ so that the target documents semantically matched to the source document $d_s$ have higher similarity scores. In practice, the documents may contain structural information like passage/ sentence/ word boundaries and different text length characteristics. The task can be formalized as a regression or classification problem depending on the type of data labels. A summary of key notations in this work is presented in Table \ref{tab:notation}. 

\begin{table}[!t]
  	\footnotesize
	\caption{A summary of key notations in this work.}
	\vspace{-0.4cm}
	\begin{tabular}
		{ p{0.07\textwidth} | p{0.38\textwidth}} \hline  \hline
		$d_s, \mathcal{D}_s$ & The source document and the set of all source documents \\\hline
		$d_c, \mathcal{D}_c$ & The candidate document and the set of all candidate documents \\\hline
		$\mathbf{E}(d_s), \mathbf{E}(d_c)$ & The learned dense vector representations for $d_s$ and $d_c$ \\ \hline
		$L_1, L_2$  & The number of layers in the sentence level Transformers and in the document level Transformers\\ \hline
		$S_i, \mathbf{E}(S_i)$ & The i-th sentence block in the document and the sequence of word representations for $S_i$\\ \hline
		$W^i_j$ & The j-th word in the i-th sentence block in the document \\ \hline
		$L_d, L_s$ & The length of a document by sentence blocks and the length of a sentence block by words \\ \hline
		$t(W^i_j)$ & The token embedding for $W^i_j$ \\ \hline
		$p(W^i_j)$ & The position embedding for $W^i_j$  \\ \hline
		$\mathbf{T}^i_j, \mathbf{S}_i$ & The contextual token representation learned by sentence level Transformers for $W^i_j$ \ and the contextual sentence block representation learned by document level Trnasformers for $S_i$ \\ \hline 
		$b$, $H$, $A$, $L$& The batch size, the hidden size, number of attention heads and layers in Transformers \\ \hline \hline
	\end{tabular}\label{tab:notation}
	\vspace{-10pt}
\end{table}


\subsection{Document Matching with Transformers}
\label{sec:method_transformers}


The original BERT model proposed by Devlin et. al. \cite{devlin2018bert} supports text matching as the sentence pair classification task. Two input text sequences will be concatenated and separated by a special token [SEP] to feed into the BERT model, in order to learn the contextual representation of each token. Then the contextual representation of the first token, which is the added [CLS] token, will be projected into a probability distribution estimation over different label dimensions to compute the cross-entropy loss. Directly applying this ``Single Encoder'' BERT model to the document matching task will cause two problems: 1) The input text length for each document will be very limited. On average, we can only feed at most 256 tokens per document into the BERT model to run the model fine-tuning or inference process of document matching. 2) The ``Single Encoder'' BERT model cannot be served for applications requiring high inference speed. To solve this problem, we learn query independent document representations and index them offline to serve with efficient similarity search libraries~\cite{JDH17,NIPS2017_7157}. Offline indexing of document representations requires generating dense vector representations for the two documents independently without expensive interactions in the earlier stage. This motivates us to focus on designing ``Dual Encoder'' BERT model with a Siamese network architecture, where each tower is to encode one document separately. The two towers can share model parameters. In the following sections, we introduce a basic Siamese matching model with Transformers MatchBERT (Section \ref{sec:method_match_bert}) and the Siamese hierarchical matching model SMITH (Section \ref{sec:method_smith}).
\subsubsection{\textbf{MatchBERT: A Basic Siamese Matching Model with Transformers}}
\label{sec:method_match_bert}
\begin{figure}[th]
	\center
	\includegraphics*[viewport=0mm 0mm 180mm 92mm, scale=0.32]{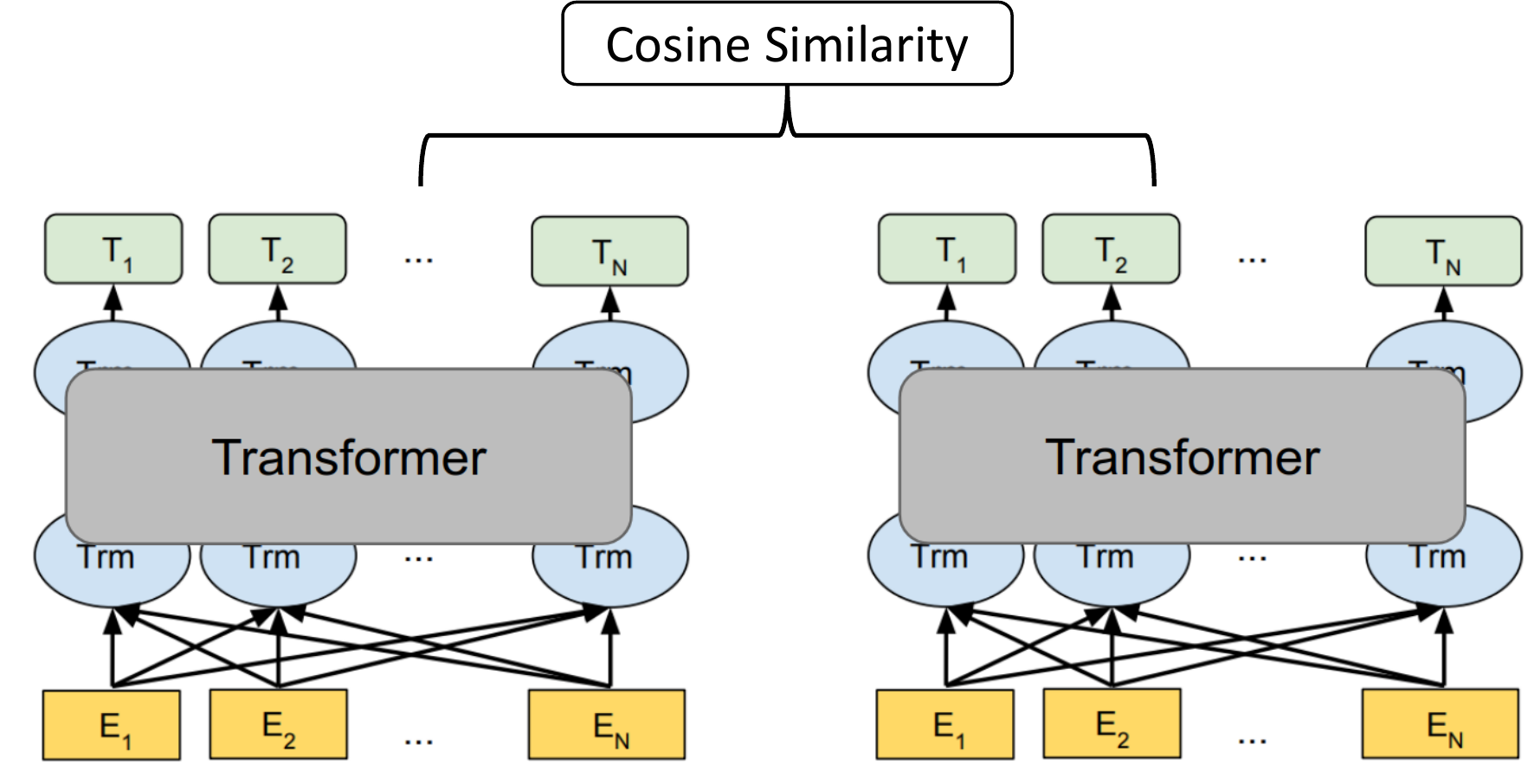}
	\vspace{-0.5cm}
	\caption{The architecture of the MatchBERT model.}\label{fig:match-bert-model}
	\vspace{-0.2cm}
\end{figure}

\begin{figure*}[th]
	\center
	\includegraphics*[viewport=0mm 0mm 320mm 125mm, scale=0.50]{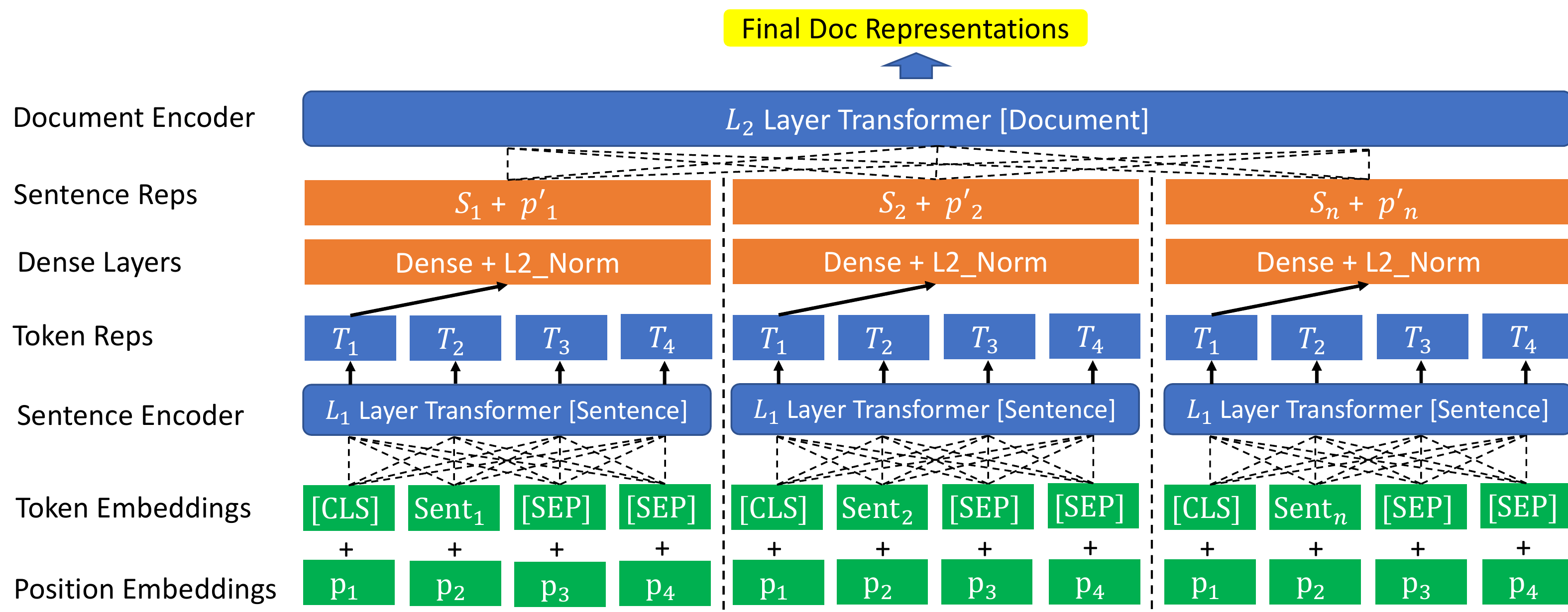}
	\vspace{-0.4cm}
	\caption{The architecture of the Multi-depth Transformer-based Hierarchical Encoder in the SMITH model for document representation learning and matching. We visualize the sentence level Transformer encoder for the 1st, 2nd and the last sentence block in a document. The output sentence representations of sentence encoders become the inputs of the document level Transformer encoder.}\label{fig:smith-model}
\end{figure*}

Figure \ref{fig:match-bert-model} shows the architecture of MatchBERT model for text matching adapted from the BERT model proposed by Devlin et. al. \cite{devlin2018bert}. There are two text encoders in MatchBERT, where each encoder is a BERT model to learn the representation of the source document $d_s$ or the candidate document $d_c$. Then we compute the cosine similarity between the pooled sequence output of two documents $cos(\mathbf{E}(d_s), \mathbf{E}(d_c))$. The text matching loss is the cross-entropy loss when we compare the document pair similarity scores with document pair labels. To handle long document input, MatchBERT will only model the first $N$ tokens of each document, where the max value of $N$ can be 512. To train the MatchBERT model, we initialize the model parameters with the open source pre-trained BERT checkpoints \footnote{\url{https://github.com/google-research/bert}} and then fine tune the model with the text matching loss. MatchBERT will be a strong baseline model.

\subsubsection{\textbf{SMITH: Siamese Hierarchical Matching Model with Transformers}}
\label{sec:method_smith}


The SMITH model, which refers to \textbf{S}iamese \textbf{M}ult\textbf{I}-depth \textbf{T}ransformer-based \textbf{H}ierarchical Encoder, is an extension of the MatchBERT model. It also adopts a Siamese network architecture, where each tower is a transformer-based hierarchical encoder to learn representations in different levels like sentence and document level of long documents. In this way, it combines the advantages of long distance dependency modeling of self-attention in Transformer encoders and hierarchical document structure modeling for long text representation learning. Figure \ref{fig:smith-model} shows the Transformer-based hierarchical encoder in the SMITH model. 


\section{Siamese Hierarchical Matching Model with Transformers}

\subsection{Hierarchical Modeling for Document Representation}
\label{sec:doc_representation}


\begin{figure}[th]
	\center
	\includegraphics*[viewport=0mm 0mm 140mm 52mm, scale=0.54]{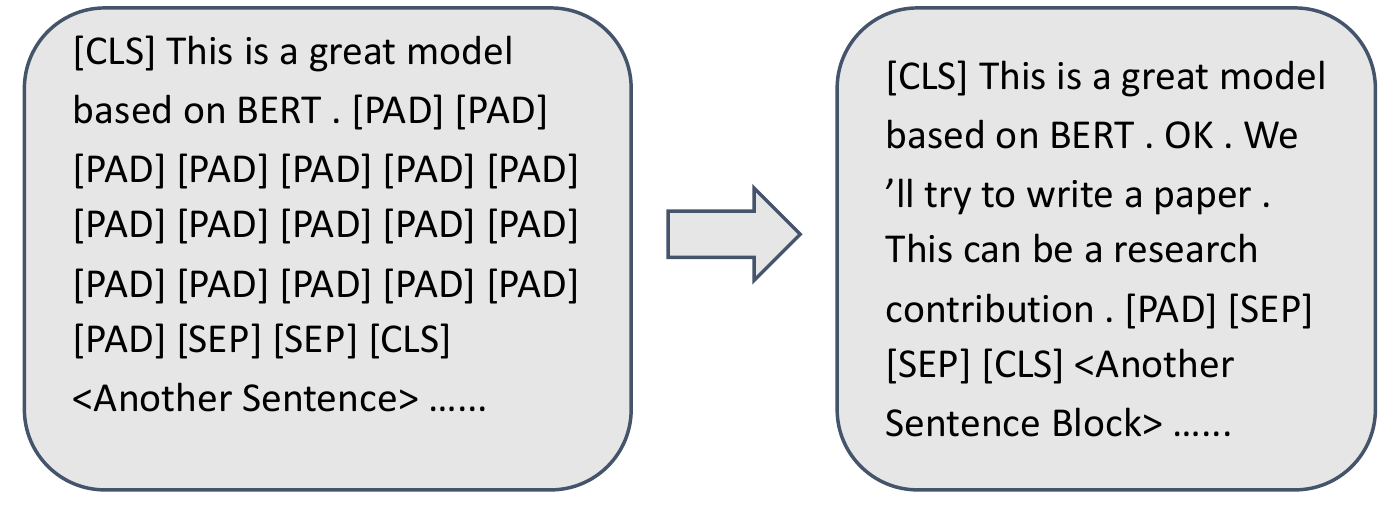}
	\vspace{-0.4cm}
	\caption{An example of splitting a document into different sentence blocks using the greedy sentence filling method. 
	Left: natural sentence splitting. Right: greedy sentence filling splitting.}\label{fig:greedy-sent-filling}
\end{figure}

\subsubsection{\textbf{Splitting Documents with Greedy Sentence Filling}}
In order to represent one document with hierarchical Transformers, we need to split the document into multiple smaller text units. A natural way to perform this step is to split a document into sentences with some off-the-shelf sentence boundary detection libraries. However, as sentence length varies a lot, padding all sentences to the same length will introduce a lot of padded tokens for short sentences, which will make the usage of the model capacity unnecessarily inefficient. We want to  preserve each sentence's semantics so methods which may break down sentences like fixed length sliding window\cite{10.1145/3331184.3331303} are not good options either. We propose a ``greedy sentence filling'' method to reduce the number of padded words and increase the actual text length the model can take as its input. Specifically, we split a document into multiple sentence blocks of predefined length so that each sentence block can contain one or more natural sentences. We try to fill as many sentences as possible into one sentence block until the block reaches the predefined max block length. When the last sentence cannot fill in the current block, we move it to the next block. When an individual sentence alone is longer than the max sentence block length, we truncate it to fit in the current block. Figure \ref{fig:greedy-sent-filling} shows an example of how we split a document into sentence blocks. We can see that greedy sentence filling greatly reduces the number of padded tokens given a fixed maximum sentence block length. 



\subsubsection{\textbf{Hierarchical Document Representation Learning}}
Let $\mathcal{D}$ denote the input document. With greedy sentence filling, we split it into a sequence of sentence blocks $\{S_1, S_2, ..., S_{L_d}\}$, where $S_i = \{W^i_1, W^i_2, ..., W^i_{L_s}\}$. $S_i$ is the i-th sentence block in the document. $W^i_j$ is the j-th word in the i-th sentence block. $L_d$ and $L_s$ denote the length of a document by sentence blocks and the length of a sentence block by words respectively. We learn the representation of each sentence $S_i$ with the Transformer encoder described in \cite{NIPS2017_Transformers}, which consists of the multi-head self-attention and a position-wise fully connected feed-forward network with residual connections \cite{DBLP:journals/corr/HeZRS15}. We firstly map the words in $S_i$ to a sequence of dense vector representations 

\begin{equation}\label{Eqn:word_dense_vectors}
    \small
	\mathbf{E}(S_i) = (\mathbf{e}^i_1, \mathbf{e}^i_2, ..., \mathbf{e}^i_{L_s})
\end{equation}
where $\mathbf{e}^i_j = t(W^i_j) + p(W^i_j)$ is the sum of the token embedding and position embedding of word  $W^i_j$ following the same practice in BERT. The token embedding is initialized randomly during pre-training and the position embedding follows the same setting as in Transformers \cite{NIPS2017_Transformers}. The sentence level Transformers will transform $\mathbf{E}(S_i)$ into a sequence of contextualized representations for words in the sentence block $\{\mathbf{T}^i_1, \mathbf{T}^i_2, ..., \mathbf{T}^i_{L_s} \}$. Following the setting in BERT model, we use the contextual representation of the first token, the added [CLS] token, as the learned representation of the whole sentence block. We add another dense layer and perform a L2 normalization on the sentence block representation. The final sentence block representation also adds the sentence block position embedding to model the sentence block location in the document. 


With the learned sentence block representations from the sentence level Transformers and the sentence block position embeddings, the document level Transformer encoders will produce a sequence of contextual sentence block representations  $\{\mathbf{S}_1, \mathbf{S}_2, ..., \mathbf{S}_{L_d} \}$. We still use the first contextual sentence block representation  as the representation for the whole document. There will be another dense layer added to transform the document representation with L2 normalization before we compute the cosine similarity between the two document representations in the document pair $(d_s, d_c)$.

\subsubsection{\textbf{Memory and Time Complexity Analysis of Two Level Hierarchical Transformers}}
Next let's analyze memory and time complexity for the two level hierarchical Transformers. The attention mechanism used in Transformer is the scaled dot-product attention, which performs transformation from a query and a set of key-value pairs to an output.   The output representation is defined as a weighted sum of the values, where the weight to each value is computed as the interaction score between the query and the corresponding key normalized by the softmax function. Specifically, given the input query embeddings $\mathcal{Q}$, key embeddings $\mathcal{K}$ and value embeddings $\mathcal{V}$, where $\mathcal{Q} \in \mathbb{R}^{b \times l_\mathcal{Q}\times H}$, $\mathcal{K} \in \mathbb{R}^{b \times l_\mathcal{K}\times H}$, $\mathcal{V} \in \mathbb{R}^{b \times l_\mathcal{V}\times H}$, the scaled dot-product attention is defined as:

\begin{equation}\label{Eqn:transformer_scaled_dot_prod_att}
\small
\text{Attention}(\mathcal{Q}, \mathcal{K}, \mathcal{V}) = \text{softmax}\Big( \frac{\mathcal{Q}\mathcal{K}^T}{\sqrt{d}}\Big)\mathcal{V}
\end{equation}

where $l_\mathcal{Q}$, $l_\mathcal{K}$, $l_\mathcal{V}$ are the number of tokens in each sequence and $l_\mathcal{K} = l_\mathcal{V}$. $b$ is the batch size and $H$ is the hidden size. To understand the memory cost of Transformers, we can focus on the attention computation in Equation \ref{Eqn:transformer_scaled_dot_prod_att}. Let us assume  $l_\mathcal{K} = l_\mathcal{V} =  l_\mathcal{Q} = N$, then the term $\mathcal{Q}\mathcal{K}^T$ has the shape $[b, N, N]$, where $N$ is the maximum input sequence length. Let $A$ and $L$ denote the number of attention heads and layers in Transformers, then the memory complexity of the attention computation in Transformers is $O(b \cdot A \cdot N^2 \cdot L)$. This is why the memory cost of the scaled dot-product attention in Transformers grows quadratically as the increasing of the input sequence length, which makes it difficult to directly apply Transformers to very long input sequences \footnote{This is the memory cost of Transformers without considering the feed-forward layers. For the complete memory complexity analysis results including both attention and feed-forward layers in Transformers, we refer the interested readers to \cite{kitaev2020reformer}.}. Similar conclusions also hold for the time complexity of scaled dot-product used in Transformers. For two level hierarchical Transformers, let $L_s$ denote the max sentence block length by tokens. Then we will split a  document into $\frac{N}{L_s}$ sentence blocks. The memory complexity of the attention computation of sentence/ document level Transformers is

\begin{small}
\begin{eqnarray}\label{Eqn:two_level_transformer_attention_memory_cost}
&& b \cdot A \cdot {L_s}^2 \cdot L \cdot \frac{N}{L_s} + b \times A \times (\frac{N}{L_s})^2 \cdot L \\ \nonumber
&=&  ( {L_s}^2 \cdot \frac{N}{L_s} + (\frac{N}{L_s})^2 ) \cdot b \cdot A \cdot L  \\ \nonumber
&=&  (L_s \cdot N + \frac{N^2}{L_s^2}) \cdot b \cdot A \cdot L 
\end{eqnarray}
\end{small}

Here we assume the number of attention heads and the number of layers are the same for the sentence level Transformers and document level Transformers for simplicity. Thus the memory complexity of two level hierarchical Transformers is $O(\frac{N^2}{L_s^2} \cdot b \cdot A \cdot L)$. Comparing with the original Transformers, we reduce the memory complexity by a factor of $L_s^2$ with only performing local self-attention over tokens in the same sentence block. 


%
%
%
%

\subsubsection{\textbf{Combine Representations from Different Levels}}
\label{sec:doc_combine_representation_from_different_levels}
In order to integrate learned features from different levels of document structures, we consider several settings for generating the final document representations as follows:

\textbf{Normal}: we only use the output of the document level Transformers as the final document representation.

\textbf{Sum-Concat}: we firstly compute the sum of all sentence level representations and use the concatenation of the sum with the document level representation as the final document representation.

\textbf{Mean-Concat}: we firstly compute the mean of all sentence level representations and use the concatenation of the mean with the document level representation as the final document representation.

\textbf{Attention}: we firstly compute the weighted sum of the sentence level representations with attention mechanism: $\sum_{i=1}^{L_d} \mathbf{h}_i \cdot \text{softmax}(\mathbf{h}_i \mathbf{W} \mathbf{v})$, where $\mathbf{h}_i  \in  \mathbb{R}^{H} $ is the learned representation for the i-th sentence block by the sentence level Transformers. $\mathbf{W} \in \mathbb{R}^{H \times V}$ is the projection matrix and $\mathbf{v} \in  \mathbb{R}^{V}$ is the attention model parameter. Then we concatenate the weighted sum with the document level representation as the final document representation.

\subsection{SMITH Model Pre-training}
\label{sec:smith_model_pre_training}
For the model training of SMITH, we adopt the ``pre-training + fine-tuning'' paradigm as in BERT. This approach is to firstly pre-train the model with large unlabeled  plain text in an unsupervised learning fashion, and then fine-tune the model with a supervised downstream task so that only few parameters need to be learned from scratch. In addition to the masked word language modeling task proposed by Devlin et. al. \cite{devlin2018bert}, we also propose the masked sentence block language modeling task, because one of the basic units in the SMITH encoder for modeling documents is the sentence block. Masked sentence block prediction task can help the model learn the relations between different sentence blocks and hopefully, get a better understanding of whole documents. Our pre-training loss is a sum of masked word prediction loss and the masked sentence block prediction loss. For the details of the masked word prediction task, please refer to Devlin et. al. \cite{devlin2018bert}. For the masked sentence block prediction task, we perform dynamic sentence block masking and masked sentence block prediction as follows.

\subsubsection{\textbf{Dynamic Sentence Block Masking.}}
Let $\mathbf{D} = \{\mathbf{h}_1, \mathbf{h}_2, ..., \mathbf{h}_{L_d}\}$ denote a sequence of sentence block representations learned by the sentence level Transformers.  For each document in the current batch, we randomly sample $m$ sentence blocks $\mathcal{M} = \{\mathbf{h}_k | \mathbf{h}_k \in \mathbb{R}^H, k \in \mathcal{K}\}$ and replace these sentence blocks with a randomly initialized masked sentence block vector $\hat{\mathbf{h}} \in \mathbb{R}^H$. For example, if we randomly select the 3rd and 5th sentence block for masking, the masked document becomes $\mathbf{\hat{D}} = \{\mathbf{h}_1, \mathbf{h}_2,  \hat{\mathbf{h}}, \mathbf{h}_4, \hat{\mathbf{h}},\mathbf{h}_6, ..., \mathbf{h}_{L_d}\}$. This dynamic sampling process repeats for every document in a batch in each step, so that the same document may get different masked sentence block positions in different steps. The dynamic masking strategy enables the model to predict a larger range of sentence blocks in a document compared with the opposite static masking.
 
\subsubsection{\textbf{Masked Sentence Block Prediction.}}
To perform masked sentence block prediction, we consider a multi-class sentence block classification setting similar to the masked word prediction. However, we do not have a global vocabulary for different sentence blocks. \footnote{In fact, the number of all unique sentence blocks can be unlimited considering different composition of words into sentence blocks.} Instead, we collect all the masked sentence blocks in a batch as a candidate sentence block pool, from which the model will try to predict the correct sentence block. For each masked sentence block position, the original sentence block in the current position is the positive example. The other co-masked sentence blocks in the current document and in the other documents of the same batch are the negative examples. Specifically, we apply document level Transformers on the masked document  $\mathbf{\hat{D}}$ to get a sequence of contextual sentence block representations $\{\mathbf{\hat{S}}_1, \mathbf{\hat{S}}_2, ..., \mathbf{\hat{S}}_{L_d} \}$. Then $\mathbf{\hat{S}}_k$ will be used to predict the original sentence block representation $\mathbf{h}_k$. Given a batch of $B$ masked sentence blocks with the predicted sentence block representation $\mathbf{\hat{S}} \in \mathbb{R}^{B \times H}$  and the ground truth sentence block representation  $\mathbf{h} \in \mathbb{R}^{B \times H}$ where $B = b \times m$, we can compute a pairwise similarity matrix for every masked sentence block pair in the current batch as follows:

\begin{equation}\label{Eqn:similarity_matrix_batch_pairwise}
\small
\text{Sim}(\mathbf{\hat{S}}, \mathbf{h}) = \mathbf{\hat{S}}\mathbf{h}^T
\end{equation}

where $\text{Sim}(\mathbf{\hat{S}}, \mathbf{h}) \in \mathbb{R}^{B \times B}$ where $\text{Sim}(\mathbf{\hat{S}}_j, \mathbf{h}_i)$ is the predicted similarity between the j-th predicted sentence block representation and the i-th sentence block class. We normalize it with a softmax function to transform it to the predicted probability for the the i-th sentence block class as follows:

\begin{equation}\label{Eqn:predict_probability}
\small
p(\mathbf{h}_i | \mathbf{\hat{S}}_j) = \frac{\exp( \text{Sim}(\mathbf{\hat{S}}_j, \mathbf{h}_i) )}{\sum_{r=1}^{B} \exp ( \text{Sim}(\mathbf{\hat{S}}_j, \mathbf{h}_r) )}
\end{equation}

Thus all the sentence blocks $\{\mathbf{h}_r\}$, where $r \in [1, B], r \neq j$ can be treated as randomly sampled negative classes for $ \mathbf{\hat{S}}_j$. Finally we can compute the cross-entropy loss over all masked sentence blocks and the pre-training joint loss:

\begin{equation}\label{eqn:sentence-loss}
\small
\begin{aligned}
\mathcal{L}_{\text{sp}} = - \frac{1}{B}\sum_{i=1}^B\sum_{j=1}^B\mathds{1}\{j=i\} &\log{ p(\mathbf{h}_i | \mathbf{\hat{S}}_j) } 
\end{aligned}
\end{equation}

\begin{equation}\label{eqn:pretrain-loss}
\small
\begin{aligned}
\mathcal{L}_{\text{pretrain}} = \mathcal{L}_{\text{sp}} + \mathcal{L}_{\text{wp}}  
\end{aligned}
\end{equation}

 where $\mathcal{L}_{\text{sp}} $ and $\mathcal{L}_{\text{wp}}  $ denote the masked sentence block prediction loss and the masked word prediction loss respectively.

 
%
%
\subsection{SMITH Model Fine-tuning and Inference}
\label{sec:smith_model_fine_tuning}

After model pre-training, we fine-tune SMITH model on the downstream document matching task with only the binary cross-entropy loss between the estimated matching probability and the ground truth matching label. Note that the word level and sentence block level language modeling masks added during the pre-training stage need to be removed during the fine-tuning stage to avoid losing document content information. After model pre-training and fine-tuning, the trained SMITH model can be used for document representation inference. The document representations inferred by SMITH model offline can be served online with fast similarity search libraries \cite{JDH17,NIPS2017_7157}. 


%% file: exp_part1.tex
\section{Experiments}
\label{sec:exps}
\subsection{Dataset Description}
\label{sec:data_desc}

\begin{table}[t]
    \small
	\centering
	\caption{The statistics of experimental datasets. \# of DocPairs denotes the number of document pairs. AvgSPerD, AvgWPerD, AvgWPerS denote the average number of sentences per document, average number of words per document and average number of words per sentence respectively.}
	\vspace{-0.1in}
	\label{tab:exp_data_stat_train_valid_test}
	\begin{tabular}{l|l|l|l|l|l|l}
		\hline  \hline
		Data & \multicolumn{3}{c|}{Wiki65K}          & \multicolumn{3}{c}{AAN104K}          \\ \hline
		Items                       & Train       & Valid       & Test       & Train       & Valid       & Test       \\ \hline
		\# of DocPairs              & 65,948      & 8,166      & 8,130      & 104,371     & 12,818     & 12,696     \\ \hline
		AvgSPerD    & 92.4       & 92.0      & 91.0      & 111.6      & 111.4     & 111.1     \\ \hline
		AvgWPerD      & 2035.3     & 2041.7    & 1992.3    & 3270.1     & 3251.2    & 3265.9    \\ \hline
		AvgWPerS  & 22.0       & 22.2      & 21.9      & 29.3          & 29.2         & 29.4         \\ \hline \hline
	\end{tabular}
\end{table}

Following \cite{10.1145/3308558.3313707}, we evaluate our proposed methods with two datasets: Wikipedia relevant document recommendation data (Wiki65K) and ACL Anthology Network paper citation suggestion data (AAN104K) from the previous related work. The statistics of the experimental datasets are shown in Table \ref{tab:exp_data_stat_train_valid_test}. Note that we do not use any TREC datasets like TREC 2019 Deep Learning Track data~\cite{craswell2020overview}. This is because these datasets focus on short-to-long matching like document ranking, aiming to match a short query to a set of documents, whereas our task is more on long-to-long document matching.


\subsubsection{\textbf{Relevant Document Recommendation Data}}
Relevant document recommendation can be useful in many real word applications such as news recommendation, related Web page recommendation, related QA posts recommendation, etc. We use the Wikipedia relevant document recommendation data from Jiang et. al. \cite{10.1145/3308558.3313707} as the evaluation set. The ground truth of document similarity is constructed based on the Jaccard similarity between the outgoing links of two Wikipedia documents with the assumption that similar documents have similar sets of outgoing links. The document pairs with similarities greater than 0.5 are considered as positive examples. For each positive document pair, the document with the lexicographical smaller URL is defined as the source document of the pair. Then a mismatched document from the outgoing links of the source document is sampled to generate a negative document pair. This negative sampling approach is better than random sampling from the entire corpus, since random sampled documents may be too irrelevant to make the task challenging enough to evaluate the performance of different methods. For more details of this dataset, we refer interested readers to  \cite{10.1145/3308558.3313707}. 

Note that there are around six thousands training examples and less than one thousand validation/testing examples in the Wikipedia data used in  \cite{10.1145/3308558.3313707} \footnote{\url{https://research.google/pubs/pub47856/}}. We produce a Wikipedia document matching dataset of ten times larger size as shown in the data statistics in Table \ref{tab:exp_data_stat_train_valid_test}. Thus the training/validation/testing data partition and statistics are different from the data in  \cite{10.1145/3308558.3313707} and we report the results from running the model implementation on these datasets, which are different from the numbers reported in \cite{10.1145/3308558.3313707}.  We refer to this data as Wiki65K since there are 65K training document pairs. 

\subsubsection{\textbf{Paper Citation Suggestion Data}}
Paper citation suggestion can help researchers find related works and finish paper writing in a more efficient way. Given the content of a research paper and the other paper as a candidate citation, we would like to predict whether the candidate should be cited by the paper. We use the ACL Anthology Networks (AAN) Corpus \cite{10.5555/1699750.1699759} processed by Jiang et. al. \cite{10.1145/3308558.3313707}  for the evaluation. The AAN data consists of 23,766 papers written by 18,862 authors in 373 venues related to NLP. For each paper with available text, the paper with each of its cited paper in the corpus is treated as a positive example. For each positive example, an irrelevant paper is randomly sampled to generate a negative example. The reference sections were removed to prevent the leakage of ground truth. The abstract sections were also removed to increase the difficulty of the task. We also filter document pairs with no section content or invalid UTF-8 text based on the processed data in \cite{10.1145/3308558.3313707} .  We refer to this data as AAN104K since there are 104K training document pairs. 
 
\subsubsection{\textbf{Unsupervised Language Model Pre-training Data}}
\label{sec:pretrain-data}

For the SMITH model pre-training, we create a randomly sampled Wikipedia collection, containing 714,800 documents with 43,532,832 sentences and 956,169,485 words. We pre-train SMITH model with unsupervised masked word and masked sentence block language modeling loss on this data and fine-tune the model on Wiki65K and AAN104K for the downstream document matching task.

\begin{table*}[t]
    \small
	\centering
	\caption{Comparison of different models over Wiki65K and AAN104K datasets. The best performance is highlighted in boldface. SMITH-WP+SP shows significant improvements over all baseline methods with p < 0.05 measured by Student's  t-test. Note that SMITH-Short with input documents with maximum length larger than 512 and MatchBERT with input documents with maximum length larger than 256 will trigger the out-of-memory (OOM) issues on TPU V3. There is $\times 2$ in the BestDocLen, which denotes the best setting of the maximum input document length, since all the compared models are ``Dual-Encoder''/ ``Siamese'' models. Note that all the compared models are for the document/ document matching task. Models designed for the query/ document matching task are not comparable.}  
	\vspace{-0.1in}
	\label{tab:main_eval_results}
	\begin{tabular}{l|l|l|l|l|l|l|l|l|l}
		\hline  \hline
		      &  Data  & \multicolumn{4}{c|}{Wiki65K}           & \multicolumn{4}{c}{AAN104K}           \\ \hline
		Method     & BestDocLen  & Accuracy & Precision & Recall & F1     & Accuracy & Precision & Recall & F1     \\ \hline \hline
		HAN (NAACL16)& 2048 $\times$ 2 & 0.8875   & 0.8571    & 0.9317 & 0.8928 & 0.8219	& 0.7895 &	0.8654 & 0.8257\\ \hline
		SMASH (WWW19)& 2048 $\times$ 2 & 0.9187   & 0.9177    & 0.9177 & 0.9177 & 0.8375	& 0.8224 & 0.8333 & 0.8278 \\ \hline
		MatchBERT  & 256 $\times$ 2 & 0.9316   & 0.9272    & 0.9366 & 0.9319 & 0.8355   & 0.8387    & 0.8201 & 0.8293 \\ \hline \hline
		SMITH-Short & 512 $\times$ 2 & 0.9415   & 0.9178    & 0.9699 & 0.9431 & 0.8212   & 0.8169    & 0.8161 & 0.8165 \\ \hline
		SMITH-NP   & 1536 $\times$ 2 & 0.9054	& 0.8911	& 0.9237	& 0.9071	& 0.7725	& 0.8106	& 0.7062	& 0.7548 \\ \hline
		SMITH-WP   &  1536 $\times$ 2   & 0.9492	& 0.9307    & 0.9707 &	0.9503 & 	0.8400 & 	0.8408 & 	0.8354 &	0.8381 \\ \hline
		SMITH-WP+SP &  1536 $\times$ 2  & \textbf{0.9585}$^\ddagger$	& \textbf{0.9466}$^\ddagger$	 & \textbf{0.9720}$^\ddagger$ &	\textbf{0.9591}$^\ddagger$ &	\textbf{0.8536}$^\ddagger$ &	\textbf{0.8431}$^\ddagger$ &	\textbf{0.8657}$^\ddagger$ &	\textbf{0.8543}$^\ddagger$ \\ \hline 
		$\Delta$ over SMASH  & NA & +4.33\%	& +3.15\%	& +5.92\% & +4.51\%	& +1.92\%	& +2.52\%	& +3.89\%	& +3.19\% \\ \hline 
		$\Delta$ over MatchBERT & NA  & +2.89\%	& +2.09\%	& +3.78\%	& +2.92\%	& +2.17\%	& +0.52\%	& +5.56\%	& +3.01\% \\ \hline \hline
	\end{tabular}
\end{table*}

\subsection{Experimental Setup}

\subsubsection{\textbf{Competing Methods.}}
We consider different types of methods for comparison as follows: 
1) \textbf{Hierarchical Attention Network (HAN).} The HAN model \cite{yang-etal-2016-hierarchical} is a hierarchical attention model to learn document representations. For each sentence in the document, it applied an attention-based RNN to learn the sentence representation and to attend differentially to more or less important document content.
2) \textbf{SMASH.} The SMASH model \cite{10.1145/3308558.3313707} is the state-of-the-art model for long document matching. It adopts a Siamese multi-depth attention-based hierarchical recurrent neural network (SMASH RNN) to learn long document representations for matching. 
3) \textbf{MatchBERT.} The MatchBERT model has been presented in Section \ref{sec:method_match_bert}. 
4) \textbf{SMITH.}  This is our proposed method. We fixed the document level Transformers as 3 layers and tried several SMITH model variations as follows:

\begin{itemize}
    \item \textbf{SMITH-Short}: the SMITH model with loading the pre-trained BERT checkpoint released by Devlin et al. \cite{devlin2018bert}. We load the BERT-Base checkpoint pre-trained on uncased text and then fine-tune the model with only the document matching loss. The maximum input text length is 512 (4 sentence blocks with 128 tokens per sentence block).
    \item \textbf{SMITH-NP}: the SMITH model without language modeling pre-training stage and trained from randomly initialized model parameters. We only train SMITH-NP model with the document matching data using text matching loss. 
    \item \textbf{SMITH-WP}: the SMITH model pre-trained with masked word prediction loss in the pre-training collection and then fine-tuned with document matching loss on the downstream matching task data. 
    \item \textbf{SMITH-WP+SP}: the SMITH model pre-trained with both masked word prediction loss and masked sentence block prediction loss on the pre-training collection and then fine-tuned with document matching loss on the downstream matching task data. 
\end{itemize}

Note that we do not compare with any interaction-focused models like DRMM~\cite{Guo:2016:DRM:2983323.2983769}, K-NRM~\cite{Xiong:2017:ENA:3077136.3080809}, Duet~\cite{Mitra:2017:LMU:3038912.3052579} or MatchPyramid~\cite{DBLP:conf/aaai/PangLGXWC16}. These models either do not scale to long documents or require heavy interactions between word pairs in two text sequences, which will lead to long inference latency in practice. Thus all the compared methods belong to representation-focused models or ``Dual-Encoder'' models where the document representation can be learned offline in advance before serving online with fast similarity search libraries \cite{JDH17,NIPS2017_7157}. Models like DRMM, KNRM, Duet, etc. are proposed for short-to-long text matching like query/document matching instead of long-to-long text matching that we focus on in this paper. 

\subsubsection{\textbf{Evaluation Methodology}.}
We formalize the document matching task as a classification task where we would like to predict whether two documents are relevant or not given a document pair. Thus we consider standard classification metrics including accuracy, precision, recall and F1-score for the evaluation.

\subsubsection{\textbf{Parameter Settings and Implementation Details}} 
All models are implemented with TensorFlow\footnote{\url{https://www.tensorflow.org/}}. We use TPU V3\footnote{https://cloud.google.com/tpu/} for the model pre-training and fine-tuning. For the model pre-training stage, we pre-train SMITH on the Wikipedia collection presented in Section \ref{sec:pretrain-data} around 68 epochs until the validation loss does not decrease significantly. Pre-training of SMITH with 2 layers in the sentence level Transformers and 3 layers in the document level Transformers with max document length 1024 takes 50 minutes for one epoch on the Wikipedia collection and the pre-training stage takes around 57 hours. The pre-training loss depends on the model variation types (masked word prediction loss for SMITH-WP or the sum of masked word prediction loss and masked sentence block prediction loss for SMITH-WP+SP). We dynamically mask 2 sentence blocks per document if we use the masked sentence block prediction loss presented in Section \ref{sec:smith_model_pre_training} during pre-training. The masked word prediction task follows the similar way by Devlin et al. \cite{devlin2018bert}. The sentence block length is 32. We tune max number of sentence blocks with values in $\{32, 48, 64\}$. Thus the max document length can be values in $\{1024, 1536, 2048\}$. We finally set the number of sentence blocks as $48$ (max document length is $1536$) for both Wiki65K and AAN104K, which achieves the best performance on the validation data. We tune parameters using the validation dataset and report model performance on the test dataset. 
Let $L_1, H_1, A_1$  and $L_2, H_2, A_2$ denote the number of Transformer layers, the hidden size, the number of attention heads in sentence level Transformers and document level Transformers. We fix $L_2 = 3$ and tune $L_1$ with values in $\{2, 4, 6\}$. We finally set $L_1=6$, $H_2=256, A_2=4$ and $H_1=256, A_1=4$ for both Wiki65K and AAN104K. Both training and evaluation batch size are 32.  We optimize the models using Adam with learning rate 5e-5, $\beta_1=0.9, \beta_2=0.999, \epsilon = 1e-6$. The dropout rate in all layers is 0.1. 

For the model fine-tuning stage, the hyper-parameters are almost the same to those used in the pre-training stage. The max number of training steps is 100K. The number of learning rate warm up steps is 10K. We remove both masked word prediction loss and masked sentence block prediction loss during fine-tuning, and update the pre-trained model parameters only using the document matching loss. The fine-tuning stage takes much less time ranging from 4 to 12 hours depending on the model and data settings.

\subsection{Evaluation Results}



We present evaluation results of different models over Wiki65K and AAN104K data  in Table \ref{tab:main_eval_results}. We summarize our observations as follows: 1) Both SMITH-WP and SMITH-WP+SP models outperform all the baseline methods including the stage-of-the-art long text matching method SMASH and MatchBERT based on pre-trained BERT models on both Wiki65K and AAN104K consistently. The comparison between SMITH-WP/ SMITH-Short and MatchBERT shows the effectiveness of introducing hierarchical document structure modeling with sentence level and document level Transformers for long document representation learning and matching. 2) If we compare the SMITH model settings with the pre-training stage (SMITH-Short, SMITH-WP, SMITH-WP+SP) with the SMITH model settings without the pre-training stage (SMITH-NP), we can find that language modeling pre-training can help increase the performance of the downstream document matching task by a large margin. Thus better language understanding via large scale language modeling pre-training will lead to better downstream task performance, which is consistent with the findings by Devlin et al. \cite{devlin2018bert}. 3) Both SMITH-WP and SMITH-WP+SP outperform SMITH-Short, which is initialized by the pre-trained open source BERT model. We think the main reason is that currently SMITH-Short can only process at most $512$ tokens due to TPU memory issues, which will hurt the performance. On the other hand, Both SMITH-WP and SMITH-WP+SP can process as long as $2048$ tokens, which is a better setting for long document representation learning. 4) If we compare SMITH-WP with SMITH-WP+SP, we can find that adding the masked sentence block prediction task presented in Section \ref{sec:smith_model_pre_training} during the pre-training stage can also be helpful to improve the downstream document matching performance. The masked word prediction task proposed  by Devlin et al. \cite{devlin2018bert} can capture the word relations and dependencies in the pre-training corpus, whereas the masked sentence block prediction task can additionally force the model to learn the sentence-level relations and dependencies. Thus combining the masked word prediction task and the masked sentence block prediction task can contribute to a better pre-training language model for long document content understanding and better downstream document matching task performance. 

%% file: exp_part2.tex
\subsection{Impact of Document Length}

\begin{figure}[t]
	\centering
	\begin{subfigure}[b]{0.24\textwidth}
		\includegraphics[width=\textwidth]{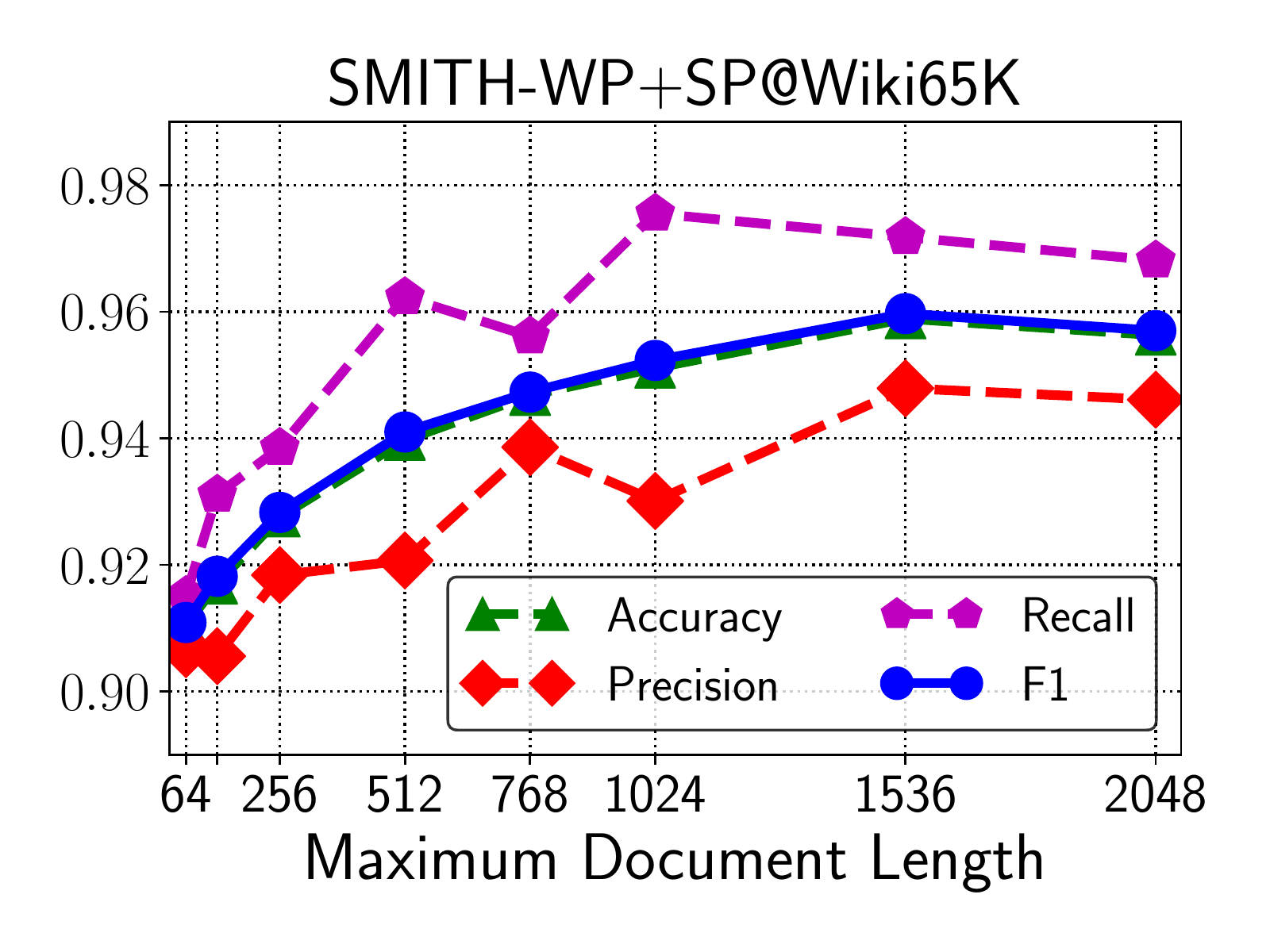}
		\label{fig:doc-len-wiki}
	\end{subfigure}
	\hspace{-0.15in}
	~ 
	\begin{subfigure}[b]{0.24\textwidth}
		\includegraphics[width=\textwidth]{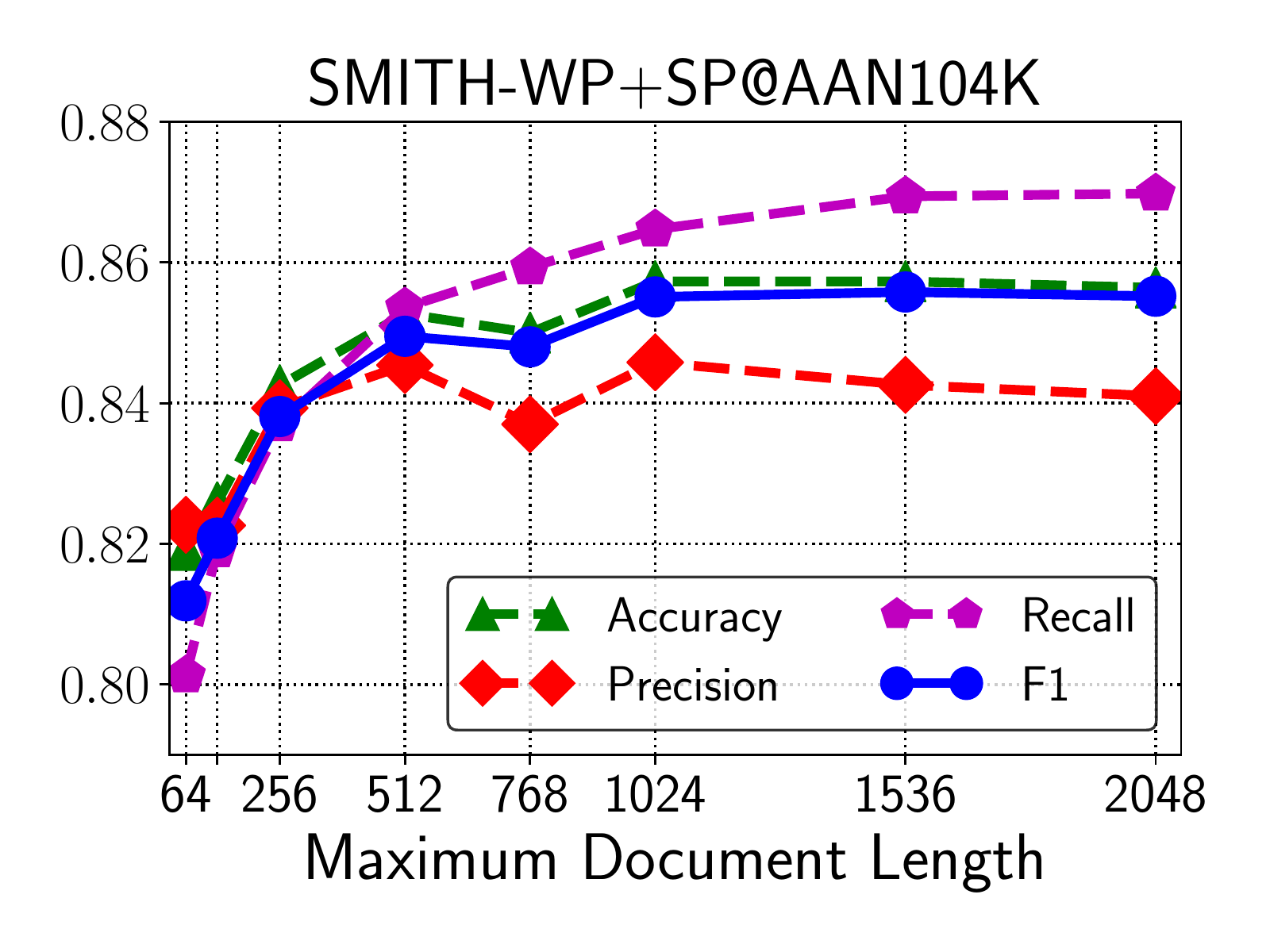}
		\label{fig:doc-len-aan}
	\end{subfigure}
	\vspace{-0.4in}
	\caption{Performance of SMITH-WP+SP on the validation data with different choices of maximum document length.}\label{fig:tune_doc_len}
\end{figure}

We further analyze the impact of document length on the document matching performance. We fix the number of layers in the sentence level and the document level Transformers as 4 and 3, the max sentence block length as 32. Then we vary the number of looped sentence blocks per document for SMITH-WP+SP on Wiki65K and AAN104K with different values from 2 to 64. Thus the maximum document length increases from 64 to 2048. The performances of SMITH-WP+SP with different choices of maximum document length is shown in Figure \ref{fig:tune_doc_len}. We can find that in general SMITH-WP+SP will achieve better performance as the maximum document length increases. This confirms the necessity of long text content modeling for document matching. The SMITH model which enjoys longer input text lengths compared with other standard self-attention models is a better choice for long document representation learning and matching. We also studied the impact of different sentence block lengths and found it has no major impact on the final performance.

%
%
%
%

\subsection{Impact of the Number of Layers in Sentence Level Transformers}
Next we analyze the impact of the number of layers in the sentence level Transformers on the final document matching performances. We set the document level Transformer layers as 3, the maximum sentence block length as 32, the number of sentence blocks per document as 48. So the maximum document length is 1536. Then we vary the number of layers in the sentence level Transformers and observe the change of the performances of SMITH-WP+SP. The results are shown in Table \ref{tab:impact_sent_level_transformer_layers}. We can find that the setting with 4 or 6 layers in the sentence level Transformer layers is slightly better than the setting with only 2 layers in the sentence level Transformers. Increasing the layers in the sentence level Transformers can help the model to learn sentence block semantic representation with more high level interactions. However, it also leads to larger memory cost such as the intermediate activation in each layer. Thus in practice this hyper-parameter has to be tuned with the validation dataset for the trade-off between the model capacity and memory cost.

\begin{table}[t]
	\small
	\centering
	\caption{The document matching performance  with different choices of the number of layers in the sentence level Transformers on the validation data. $L_1$ denotes the number of layers in the sentence level Transformers. } 
	\vspace{-0.3cm}
	\label{tab:impact_sent_level_transformer_layers}
	\begin{tabular}{l|l|l|l|l|l}
		\hline \hline
		Data                     & $L_1$ & Accuracy & Precision & Recall & F1     \\ \hline \hline
		\multirow{3}{*}{Wiki65K} & 2    & 0.9537   & 0.9449    & 0.9635 & 0.9541 \\ \cline{2-6} 
		& 4    & 0.9589   & \textbf{0.9479}    & 0.9718 & 0.9597 \\ \cline{2-6} 
		& 6    & \textbf{0.9594}   & 0.9426   & \textbf{0.9784} & \textbf{0.9602} \\ \hline
		\multirow{3}{*}{AAN104K} & 2    & 0.8566   & 0.8470    & 0.8612 & 0.8540 \\ \cline{2-6} 
		& 4    & 0.8573   & 0.8426    & \textbf{0.8694} &  \textbf{0.8558} \\ \cline{2-6} 
		& 6    & \textbf{0.8580}   & \textbf{0.8508}    & 0.8591 & 0.8549 \\ \hline \hline
	\end{tabular}
\end{table}

\subsection{Impact of Combining Representations from Different Levels}


\begin{table}[t]
\small
	\centering
	\caption{The document matching performance  with different choices of the document representation combining methods presented in Section \ref{sec:doc_combine_representation_from_different_levels} on the validation data.} 
	\vspace{-0.3cm}
	\label{tab:impact_different_doc_combine_mode}
\begin{tabular}{l|l|l|l|l|l}
\hline \hline
Data                     & Combine & Accuracy & Precision & Recall & F1     \\ \hline
\multirow{4}{*}{Wiki65K} & Normal        & \textbf{0.9594}   & \textbf{0.9426}    & 0.9784 & \textbf{0.9602} \\ \cline{2-6} 
                         & Sum-Concat   & 0.9192   & 0.9103    & 0.9301 & 0.9201 \\ \cline{2-6} 
                         & Mean-Concat  & 0.9221   & 0.8924    & 0.9599 & 0.9249 \\ \cline{2-6} 
                         & Attention     & 0.9431   & 0.9099    & \textbf{0.9836} & 0.9453 \\ \hline
\multirow{4}{*}{AAN104K} & Normal        & \textbf{0.8580}   & \textbf{0.8508}    & 0.8591 & \textbf{0.8549} \\ \cline{2-6} 
                         & Sum-Concat   & 0.7632   & 0.7924    & 0.6962 & 0.7412 \\ \cline{2-6} 
                         & Mean-Concat  & 0.7061   & 0.6387    & \textbf{0.9131} & 0.7516 \\ \cline{2-6} 
                         & Attention     & 0.8434   & 0.8162    & 0.8758 & 0.8450 \\ \hline \hline
\end{tabular}
\end{table}

As presented in Section \ref{sec:doc_combine_representation_from_different_levels}, we evaluate the performances of SMITH-WP+SP with different methods to combine representations from different levels. Table \ref{tab:impact_different_doc_combine_mode} shows the document matching performance with different choices of document representation combing methods. We can see that the ``normal'' combing method where we only use the output of the document level Transformers as the final document representation works best. For the other three methods, the ``attention'' method is better than ``sum-concat'' and ``mean-concat''. One possible reason is that the weighted combination of sentence blocks can be helpful for generating better document representations as the attention weights can encode the relative importance of different sentence blocks on representing the document content, which is why attention based combining methods work better. The document level Transformers already learn a weighted combination based on the input sentence block representation sequences, which already provide enough signals on the importance scores of different sentence blocks in a document.





%% file: conclusion.tex

\section{Conclusions and Future Work}
\label{sec:conclu}

In this paper, we propose the Siamese Multi-depth Transformer-based Hierarchical (SMITH) Encoder for document representation learning and matching, which contains several novel design choices like two level hierarchical Transformers to adapt self-attention models for long text inputs. For model pre-training, we propose the masked sentence block language modeling task in addition to the original masked word language modeling task in BERT, to capture sentence block relations within a document. The experimental results on several benchmark datasets show that our proposed SMITH model outperforms previous state-of-the-art Siamese matching models including HAN, SMASH and BERT for long-form document matching. Moreover, our proposed model increases the maximum input text length from 512 to 2048 when compared with BERT-based baseline methods. 

As a part of this work, we plan to release a large scale benchmark collection for the document matching task so that it is easier for researchers to compare different document matching methods in the future. It is also interesting to investigate how to utilize the learned document representation from Transformer-based hierarchical encoders for other document-level language understanding tasks like document classification, clustering and ranking.



%% file: CIKM2020-SMITH-Text-Match.bbl

\begin{thebibliography}{40}


\ifx \showCODEN    \undefined \def \showCODEN     #1{\unskip}     \fi
\ifx \showDOI      \undefined \def \showDOI       #1{#1}\fi
\ifx \showISBNx    \undefined \def \showISBNx     #1{\unskip}     \fi
\ifx \showISBNxiii \undefined \def \showISBNxiii  #1{\unskip}     \fi
\ifx \showISSN     \undefined \def \showISSN      #1{\unskip}     \fi
\ifx \showLCCN     \undefined \def \showLCCN      #1{\unskip}     \fi
\ifx \shownote     \undefined \def \shownote      #1{#1}          \fi
\ifx \showarticletitle \undefined \def \showarticletitle #1{#1}   \fi
\ifx \showURL      \undefined \def \showURL       {\relax}        \fi
\providecommand\bibfield[2]{#2}
\providecommand\bibinfo[2]{#2}
\providecommand\natexlab[1]{#1}
\providecommand\showeprint[2][]{arXiv:#2}

\bibitem[\protect\citeauthoryear{Bowman, Angeli, Potts, and Manning}{Bowman
  et~al\mbox{.}}{2015}]%
        {bowman-etal-2015-large}
\bibfield{author}{\bibinfo{person}{S.~R. Bowman}, \bibinfo{person}{G. Angeli},
  \bibinfo{person}{C. Potts}, {and} \bibinfo{person}{C.~D. Manning}.}
  \bibinfo{year}{2015}\natexlab{}.
\newblock \showarticletitle{A large annotated corpus for learning natural
  language inference}. In \bibinfo{booktitle}{\emph{EMNLP '15}}.
  \bibinfo{pages}{632--642}.
\newblock


\bibitem[\protect\citeauthoryear{Child, Gray, Radford, and Sutskever}{Child
  et~al\mbox{.}}{2019}]%
        {DBLP:journals/corr/abs-1904-10509}
\bibfield{author}{\bibinfo{person}{R. Child}, \bibinfo{person}{S. Gray},
  \bibinfo{person}{A. Radford}, {and} \bibinfo{person}{I. Sutskever}.}
  \bibinfo{year}{2019}\natexlab{}.
\newblock \bibinfo{title}{Generating Long Sequences with Sparse Transformers}.
\newblock
\newblock
\showeprint{1904.10509}


\bibitem[\protect\citeauthoryear{Craswell, Mitra, Yilmaz, Campos, and
  Voorhees}{Craswell et~al\mbox{.}}{2020}]%
        {craswell2020overview}
\bibfield{author}{\bibinfo{person}{N. Craswell}, \bibinfo{person}{B. Mitra},
  \bibinfo{person}{E. Yilmaz}, \bibinfo{person}{D. Campos}, {and}
  \bibinfo{person}{E.~M Voorhees}.} \bibinfo{year}{2020}\natexlab{}.
\newblock \bibinfo{title}{Overview of the {TREC} 2019 deep learning track}.
\newblock
\newblock
\showeprint{2003.07820}


\bibitem[\protect\citeauthoryear{Dai and Callan}{Dai and Callan}{2019}]%
        {10.1145/3331184.3331303}
\bibfield{author}{\bibinfo{person}{Z. Dai} {and} \bibinfo{person}{J. Callan}.}
  \bibinfo{year}{2019}\natexlab{}.
\newblock \showarticletitle{Deeper Text Understanding for IR with Contextual
  Neural Language Modeling}. In \bibinfo{booktitle}{\emph{SIGIR '19}}.
\newblock


\bibitem[\protect\citeauthoryear{Dai, Yang, Yang, Carbonell, Le, and
  Salakhutdinov}{Dai et~al\mbox{.}}{2019}]%
        {DBLP:journals/corr/abs-1901-02860}
\bibfield{author}{\bibinfo{person}{Z. Dai}, \bibinfo{person}{Z. Yang},
  \bibinfo{person}{Y. Yang}, \bibinfo{person}{J.~G. Carbonell},
  \bibinfo{person}{Q.~V. Le}, {and} \bibinfo{person}{R. Salakhutdinov}.}
  \bibinfo{year}{2019}\natexlab{}.
\newblock \bibinfo{title}{Transformer-XL: Attentive Language Models Beyond a
  Fixed-Length Context}.
\newblock
\newblock
\showeprint{1901.02860}


\bibitem[\protect\citeauthoryear{Devlin, Chang, Lee, and Toutanova}{Devlin
  et~al\mbox{.}}{2018}]%
        {devlin2018bert}
\bibfield{author}{\bibinfo{person}{J. Devlin}, \bibinfo{person}{M. Chang},
  \bibinfo{person}{K. Lee}, {and} \bibinfo{person}{K. Toutanova}.}
  \bibinfo{year}{2018}\natexlab{}.
\newblock \bibinfo{title}{BERT: Pre-training of Deep Bidirectional Transformers
  for Language Understanding}.
\newblock
\newblock
\showeprint[arxiv]{1810.04805}


\bibitem[\protect\citeauthoryear{Dolan and Brockett}{Dolan and
  Brockett}{2005}]%
        {dolan-brockett-2005-automatically}
\bibfield{author}{\bibinfo{person}{W.~B. Dolan} {and} \bibinfo{person}{C.
  Brockett}.} \bibinfo{year}{2005}\natexlab{}.
\newblock \showarticletitle{Automatically Constructing a Corpus of Sentential
  Paraphrases}. In \bibinfo{booktitle}{\emph{IWP 2005}}.
  \bibinfo{pages}{9--16}.
\newblock


\bibitem[\protect\citeauthoryear{Guo, Fan, Ai, and Croft}{Guo
  et~al\mbox{.}}{2016}]%
        {Guo:2016:DRM:2983323.2983769}
\bibfield{author}{\bibinfo{person}{J. Guo}, \bibinfo{person}{Y. Fan},
  \bibinfo{person}{Q. Ai}, {and} \bibinfo{person}{W.~B. Croft}.}
  \bibinfo{year}{2016}\natexlab{}.
\newblock \showarticletitle{A Deep Relevance Matching Model for Ad-hoc
  Retrieval}. In \bibinfo{booktitle}{\emph{CIKM '16}}. \bibinfo{pages}{55--64}.
\newblock


\bibitem[\protect\citeauthoryear{Guo, Fan, Pang, Yang, Ai, Zamani, Wu, Croft,
  and Cheng}{Guo et~al\mbox{.}}{2019}]%
        {DBLP:journals/corr/abs-1903-06902}
\bibfield{author}{\bibinfo{person}{J. Guo}, \bibinfo{person}{Y. Fan},
  \bibinfo{person}{L. Pang}, \bibinfo{person}{L. Yang}, \bibinfo{person}{Q.
  Ai}, \bibinfo{person}{H. Zamani}, \bibinfo{person}{C. Wu},
  \bibinfo{person}{W.~B. Croft}, {and} \bibinfo{person}{X. Cheng}.}
  \bibinfo{year}{2019}\natexlab{}.
\newblock \bibinfo{title}{A Deep Look into Neural Ranking Models for
  Information Retrieval}.
\newblock
\newblock
\showeprint{1903.06902}


\bibitem[\protect\citeauthoryear{He, Zhang, Ren, and Sun}{He
  et~al\mbox{.}}{2015}]%
        {DBLP:journals/corr/HeZRS15}
\bibfield{author}{\bibinfo{person}{K. He}, \bibinfo{person}{X. Zhang},
  \bibinfo{person}{S. Ren}, {and} \bibinfo{person}{J. Sun}.}
  \bibinfo{year}{2015}\natexlab{}.
\newblock \bibinfo{title}{Deep Residual Learning for Image Recognition}.
\newblock
\newblock
\showeprint{1512.03385}


\bibitem[\protect\citeauthoryear{Ho, Kalchbrenner, Weissenborn, and
  Salimans}{Ho et~al\mbox{.}}{2019}]%
        {ho2019axial}
\bibfield{author}{\bibinfo{person}{J. Ho}, \bibinfo{person}{N. Kalchbrenner},
  \bibinfo{person}{D. Weissenborn}, {and} \bibinfo{person}{T. Salimans}.}
  \bibinfo{year}{2019}\natexlab{}.
\newblock \bibinfo{title}{Axial Attention in Multidimensional Transformers}.
\newblock
\newblock
\showeprint[arxiv]{1912.12180}


\bibitem[\protect\citeauthoryear{Hu, Lu, Li, and Chen}{Hu
  et~al\mbox{.}}{2014}]%
        {DBLP:conf/nips/HuLLC14}
\bibfield{author}{\bibinfo{person}{B. Hu}, \bibinfo{person}{Z. Lu},
  \bibinfo{person}{H. Li}, {and} \bibinfo{person}{Q. Chen}.}
  \bibinfo{year}{2014}\natexlab{}.
\newblock \showarticletitle{Convolutional Neural Network Architectures for
  Matching Natural Language Sentences}. In \bibinfo{booktitle}{\emph{NIPS
  '14}}. \bibinfo{pages}{2042--2050}.
\newblock


\bibitem[\protect\citeauthoryear{Huang, He, Gao, Deng, Acero, and Heck}{Huang
  et~al\mbox{.}}{2013}]%
        {DBLP:conf/cikm/HuangHGDAH13}
\bibfield{author}{\bibinfo{person}{P. Huang}, \bibinfo{person}{X. He},
  \bibinfo{person}{J. Gao}, \bibinfo{person}{L. Deng}, \bibinfo{person}{A.
  Acero}, {and} \bibinfo{person}{L.~P. Heck}.} \bibinfo{year}{2013}\natexlab{}.
\newblock \showarticletitle{Learning Deep Structured Semantic Models for Web
  Search using Clickthrough Data}. In \bibinfo{booktitle}{\emph{CIKM '13}}.
  \bibinfo{pages}{2333--2338}.
\newblock


\bibitem[\protect\citeauthoryear{Jiang, Zhang, Li, Bendersky, Golbandi, and
  Najork}{Jiang et~al\mbox{.}}{2019}]%
        {10.1145/3308558.3313707}
\bibfield{author}{\bibinfo{person}{J. Jiang}, \bibinfo{person}{M. Zhang},
  \bibinfo{person}{C. Li}, \bibinfo{person}{M. Bendersky}, \bibinfo{person}{N.
  Golbandi}, {and} \bibinfo{person}{M. Najork}.}
  \bibinfo{year}{2019}\natexlab{}.
\newblock \showarticletitle{Semantic Text Matching for Long-Form Documents}. In
  \bibinfo{booktitle}{\emph{WWW '19}}. \bibinfo{pages}{795--806}.
\newblock


\bibitem[\protect\citeauthoryear{Johnson, Douze, and J{\'e}gou}{Johnson
  et~al\mbox{.}}{2017}]%
        {JDH17}
\bibfield{author}{\bibinfo{person}{J. Johnson}, \bibinfo{person}{M. Douze},
  {and} \bibinfo{person}{H. J{\'e}gou}.} \bibinfo{year}{2017}\natexlab{}.
\newblock \bibinfo{title}{Billion-scale similarity search with GPUs}.
\newblock
\newblock
\showeprint{1702.08734}


\bibitem[\protect\citeauthoryear{Kitaev, Kaiser, and Levskaya}{Kitaev
  et~al\mbox{.}}{2020}]%
        {kitaev2020reformer}
\bibfield{author}{\bibinfo{person}{N. Kitaev}, \bibinfo{person}{L. Kaiser},
  {and} \bibinfo{person}{A. Levskaya}.} \bibinfo{year}{2020}\natexlab{}.
\newblock \showarticletitle{Reformer: The Efficient Transformer}.
\newblock In \bibinfo{booktitle}{\emph{ICLR '20}}.
\newblock


\bibitem[\protect\citeauthoryear{Li and Xu}{Li and Xu}{2014}]%
        {10.5555/2683840}
\bibfield{author}{\bibinfo{person}{H. Li} {and} \bibinfo{person}{J. Xu}.}
  \bibinfo{year}{2014}\natexlab{}.
\newblock \bibinfo{booktitle}{\emph{Semantic Matching in Search}}.
\newblock \bibinfo{publisher}{Now Publishers Inc.}, \bibinfo{address}{Hanover,
  MA, USA}.
\newblock
\showISBNx{1601988044}


\bibitem[\protect\citeauthoryear{Lowe, Pow, Serban, and Pineau}{Lowe
  et~al\mbox{.}}{2015}]%
        {DBLP:journals/corr/LowePSP15}
\bibfield{author}{\bibinfo{person}{R. Lowe}, \bibinfo{person}{N. Pow},
  \bibinfo{person}{I. Serban}, {and} \bibinfo{person}{J. Pineau}.}
  \bibinfo{year}{2015}\natexlab{}.
\newblock \bibinfo{title}{The Ubuntu Dialogue Corpus: {A} Large Dataset for
  Research in Unstructured Multi-Turn Dialogue Systems}.
\newblock
\newblock
\showeprint{1506.08909}


\bibitem[\protect\citeauthoryear{Mikolov, Sutskever, Chen, Corrado, and
  Dean}{Mikolov et~al\mbox{.}}{2013}]%
        {DBLP:conf/nips/MikolovSCCD13}
\bibfield{author}{\bibinfo{person}{T. Mikolov}, \bibinfo{person}{I. Sutskever},
  \bibinfo{person}{K. Chen}, \bibinfo{person}{G.~S. Corrado}, {and}
  \bibinfo{person}{J. Dean}.} \bibinfo{year}{2013}\natexlab{}.
\newblock \showarticletitle{Distributed Representations of Words and Phrases
  and their Compositionality}. In \bibinfo{booktitle}{\emph{NIPS '13}}.
  \bibinfo{pages}{3111--3119}.
\newblock


\bibitem[\protect\citeauthoryear{Mitra, Diaz, and Craswell}{Mitra
  et~al\mbox{.}}{2017}]%
        {Mitra:2017:LMU:3038912.3052579}
\bibfield{author}{\bibinfo{person}{B. Mitra}, \bibinfo{person}{F. Diaz}, {and}
  \bibinfo{person}{N. Craswell}.} \bibinfo{year}{2017}\natexlab{}.
\newblock \showarticletitle{Learning to Match Using Local and Distributed
  Representations of Text for Web Search}. In \bibinfo{booktitle}{\emph{WWW
  '17}}. \bibinfo{pages}{1291--1299}.
\newblock


\bibitem[\protect\citeauthoryear{Ounis, MacDonald, and Soboroff}{Ounis
  et~al\mbox{.}}{2008}]%
        {DBLP:conf/trec/OunisMS08}
\bibfield{author}{\bibinfo{person}{I. Ounis}, \bibinfo{person}{C. MacDonald},
  {and} \bibinfo{person}{I. Soboroff}.} \bibinfo{year}{2008}\natexlab{}.
\newblock \showarticletitle{Overview of the {TREC} 2008 Blog Track}. In
  \bibinfo{booktitle}{\emph{TREC '08}}.
\newblock


\bibitem[\protect\citeauthoryear{Pang, Lan, Guo, Xu, Wan, and Cheng}{Pang
  et~al\mbox{.}}{2016}]%
        {DBLP:conf/aaai/PangLGXWC16}
\bibfield{author}{\bibinfo{person}{L. Pang}, \bibinfo{person}{Y. Lan},
  \bibinfo{person}{J. Guo}, \bibinfo{person}{J. Xu}, \bibinfo{person}{S. Wan},
  {and} \bibinfo{person}{X. Cheng}.} \bibinfo{year}{2016}\natexlab{}.
\newblock \showarticletitle{Text Matching as Image Recognition}. In
  \bibinfo{booktitle}{\emph{AAAI '16}}. \bibinfo{pages}{2793--2799}.
\newblock


\bibitem[\protect\citeauthoryear{Peters, Neumann, Iyyer, Gardner, Clark, Lee,
  and Zettlemoyer}{Peters et~al\mbox{.}}{2018}]%
        {DBLP:journals/corr/abs-1802-05365}
\bibfield{author}{\bibinfo{person}{M.~E. Peters}, \bibinfo{person}{M. Neumann},
  \bibinfo{person}{M. Iyyer}, \bibinfo{person}{M. Gardner}, \bibinfo{person}{C.
  Clark}, \bibinfo{person}{K. Lee}, {and} \bibinfo{person}{L. Zettlemoyer}.}
  \bibinfo{year}{2018}\natexlab{}.
\newblock \bibinfo{title}{Deep contextualized word representations}.
\newblock
\newblock
\showeprint{1802.05365}


\bibitem[\protect\citeauthoryear{Qiu, Ma, Levy, Yih, Wang, and Tang}{Qiu
  et~al\mbox{.}}{2019}]%
        {qiu2019blockwise}
\bibfield{author}{\bibinfo{person}{J. Qiu}, \bibinfo{person}{H. Ma},
  \bibinfo{person}{O. Levy}, \bibinfo{person}{S.~W. Yih}, \bibinfo{person}{S.
  Wang}, {and} \bibinfo{person}{J. Tang}.} \bibinfo{year}{2019}\natexlab{}.
\newblock \bibinfo{title}{Blockwise Self-Attention for Long Document
  Understanding}.
\newblock
\newblock
\showeprint[arxiv]{1911.02972}


\bibitem[\protect\citeauthoryear{Radev, Muthukrishnan, and Qazvinian}{Radev
  et~al\mbox{.}}{2009}]%
        {10.5555/1699750.1699759}
\bibfield{author}{\bibinfo{person}{D.~R. Radev}, \bibinfo{person}{P.
  Muthukrishnan}, {and} \bibinfo{person}{V. Qazvinian}.}
  \bibinfo{year}{2009}\natexlab{}.
\newblock \showarticletitle{The ACL Anthology Network Corpus}. In
  \bibinfo{booktitle}{\emph{NLPIR4DL ’09}}. \bibinfo{pages}{54–61}.
\newblock


\bibitem[\protect\citeauthoryear{Radford}{Radford}{2018}]%
        {Radford2018ImprovingLU}
\bibfield{author}{\bibinfo{person}{A. Radford}.}
  \bibinfo{year}{2018}\natexlab{}.
\newblock \bibinfo{title}{Improving Language Understanding by Generative
  Pre-Training}.
\newblock \bibinfo{howpublished}{Preprint, OpenAI}.
\newblock


\bibitem[\protect\citeauthoryear{Rae, Potapenko, Jayakumar, and Lillicrap}{Rae
  et~al\mbox{.}}{2019}]%
        {rae2019compressive}
\bibfield{author}{\bibinfo{person}{J.~W. Rae}, \bibinfo{person}{A. Potapenko},
  \bibinfo{person}{S.~M. Jayakumar}, {and} \bibinfo{person}{T.~P. Lillicrap}.}
  \bibinfo{year}{2019}\natexlab{}.
\newblock \bibinfo{title}{Compressive Transformers for Long-Range Sequence
  Modelling}.
\newblock
\newblock
\showeprint[arxiv]{1911.05507}


\bibitem[\protect\citeauthoryear{Roy, Saffar, Grangier, and Vaswani}{Roy
  et~al\mbox{.}}{2020}]%
        {roy2020efficient}
\bibfield{author}{\bibinfo{person}{A. Roy}, \bibinfo{person}{M.~T. Saffar},
  \bibinfo{person}{D. Grangier}, {and} \bibinfo{person}{A. Vaswani}.}
  \bibinfo{year}{2020}\natexlab{}.
\newblock \bibinfo{title}{Efficient Content-Based Sparse Attention with Routing
  Transformers}.
\newblock
\newblock
\showeprint{2003.05997}


\bibitem[\protect\citeauthoryear{Sukhbaatar, Grave, Bojanowski, and
  Joulin}{Sukhbaatar et~al\mbox{.}}{2019}]%
        {DBLP:journals/corr/abs-1905-07799}
\bibfield{author}{\bibinfo{person}{S. Sukhbaatar}, \bibinfo{person}{E. Grave},
  \bibinfo{person}{P. Bojanowski}, {and} \bibinfo{person}{A. Joulin}.}
  \bibinfo{year}{2019}\natexlab{}.
\newblock \bibinfo{title}{Adaptive Attention Span in Transformers}.
\newblock
\newblock
\showeprint{1905.07799}


\bibitem[\protect\citeauthoryear{Vaswani, Shazeer, Parmar, Uszkoreit, Jones,
  Gomez, Kaiser, and Polosukhin}{Vaswani et~al\mbox{.}}{2017}]%
        {NIPS2017_Transformers}
\bibfield{author}{\bibinfo{person}{A. Vaswani}, \bibinfo{person}{N. Shazeer},
  \bibinfo{person}{N. Parmar}, \bibinfo{person}{J. Uszkoreit},
  \bibinfo{person}{L. Jones}, \bibinfo{person}{A.~N. Gomez},
  \bibinfo{person}{\L~. Kaiser}, {and} \bibinfo{person}{I. Polosukhin}.}
  \bibinfo{year}{2017}\natexlab{}.
\newblock \showarticletitle{Attention is All You Need}.
\newblock In \bibinfo{booktitle}{\emph{NIPS '17}}.
\newblock


\bibitem[\protect\citeauthoryear{Wu, Guo, Suresh, Kumar, Holtmann-Rice, Simcha,
  and Yu}{Wu et~al\mbox{.}}{2017a}]%
        {NIPS2017_7157}
\bibfield{author}{\bibinfo{person}{X. Wu}, \bibinfo{person}{R. Guo},
  \bibinfo{person}{A. Suresh}, \bibinfo{person}{S. Kumar}, \bibinfo{person}{D.
  Holtmann-Rice}, \bibinfo{person}{D. Simcha}, {and} \bibinfo{person}{F. Yu}.}
  \bibinfo{year}{2017}\natexlab{a}.
\newblock \showarticletitle{Multiscale Quantization for Fast Similarity
  Search}. In \bibinfo{booktitle}{\emph{NIPS '17}}.
  \bibinfo{pages}{5745--5755}.
\newblock


\bibitem[\protect\citeauthoryear{Wu, Wu, Xing, Zhou, and Li}{Wu
  et~al\mbox{.}}{2017b}]%
        {DBLP:conf/acl/WuWXZL17}
\bibfield{author}{\bibinfo{person}{Y. Wu}, \bibinfo{person}{W. Wu},
  \bibinfo{person}{C. Xing}, \bibinfo{person}{M. Zhou}, {and}
  \bibinfo{person}{Z. Li}.} \bibinfo{year}{2017}\natexlab{b}.
\newblock \showarticletitle{Sequential Matching Network: {A} New Architecture
  for Multi-turn Response Selection in Retrieval-Based Chatbots}. In
  \bibinfo{booktitle}{\emph{ACL '17}}. \bibinfo{pages}{163--197}.
\newblock


\bibitem[\protect\citeauthoryear{Xiong, Dai, Callan, Liu, and Power}{Xiong
  et~al\mbox{.}}{2017}]%
        {Xiong:2017:ENA:3077136.3080809}
\bibfield{author}{\bibinfo{person}{C. Xiong}, \bibinfo{person}{Z. Dai},
  \bibinfo{person}{J. Callan}, \bibinfo{person}{Z. Liu}, {and}
  \bibinfo{person}{R. Power}.} \bibinfo{year}{2017}\natexlab{}.
\newblock \showarticletitle{End-to-End Neural Ad-hoc Ranking with Kernel
  Pooling}. In \bibinfo{booktitle}{\emph{SIGIR '17}}. \bibinfo{pages}{55--64}.
\newblock


\bibitem[\protect\citeauthoryear{Yang, Ai, Guo, and Croft}{Yang
  et~al\mbox{.}}{2016a}]%
        {Yang:2016:ARS:2983323.2983818}
\bibfield{author}{\bibinfo{person}{L. Yang}, \bibinfo{person}{Q. Ai},
  \bibinfo{person}{J. Guo}, {and} \bibinfo{person}{W.~B. Croft}.}
  \bibinfo{year}{2016}\natexlab{a}.
\newblock \showarticletitle{aNMM: Ranking Short Answer Texts with
  Attention-Based Neural Matching Model}. In \bibinfo{booktitle}{\emph{CIKM
  '16}}. \bibinfo{pages}{287--296}.
\newblock


\bibitem[\protect\citeauthoryear{Yang, Yih, and Meek}{Yang
  et~al\mbox{.}}{2015}]%
        {yang-etal-2015-wikiqa}
\bibfield{author}{\bibinfo{person}{Y. Yang}, \bibinfo{person}{W. Yih}, {and}
  \bibinfo{person}{C. Meek}.} \bibinfo{year}{2015}\natexlab{}.
\newblock \showarticletitle{{W}iki{QA}: A Challenge Dataset for Open-Domain
  Question Answering}. In \bibinfo{booktitle}{\emph{EMNLP '15}}.
  \bibinfo{pages}{2013--2018}.
\newblock


\bibitem[\protect\citeauthoryear{Yang, Dai, Yang, Carbonell, Salakhutdinov, and
  Le}{Yang et~al\mbox{.}}{2019}]%
        {DBLP:journals/corr/abs-1906-08237}
\bibfield{author}{\bibinfo{person}{Z. Yang}, \bibinfo{person}{Z. Dai},
  \bibinfo{person}{Y. Yang}, \bibinfo{person}{J.~G. Carbonell},
  \bibinfo{person}{R. Salakhutdinov}, {and} \bibinfo{person}{Q.~V. Le}.}
  \bibinfo{year}{2019}\natexlab{}.
\newblock \bibinfo{title}{XLNet: Generalized Autoregressive Pretraining for
  Language Understanding}.
\newblock
\newblock
\showeprint{1906.08237}


\bibitem[\protect\citeauthoryear{Yang, Yang, Dyer, He, Smola, and Hovy}{Yang
  et~al\mbox{.}}{2016b}]%
        {yang-etal-2016-hierarchical}
\bibfield{author}{\bibinfo{person}{Z. Yang}, \bibinfo{person}{D. Yang},
  \bibinfo{person}{C. Dyer}, \bibinfo{person}{X. He}, \bibinfo{person}{A.
  Smola}, {and} \bibinfo{person}{E. Hovy}.} \bibinfo{year}{2016}\natexlab{b}.
\newblock \showarticletitle{Hierarchical Attention Networks for Document
  Classification}. In \bibinfo{booktitle}{\emph{NAACL '16}}.
  \bibinfo{pages}{1480--1489}.
\newblock


\bibitem[\protect\citeauthoryear{Yin and Sch{\"{u}}tze}{Yin and
  Sch{\"{u}}tze}{2015}]%
        {DBLP:conf/naacl/YinS15}
\bibfield{author}{\bibinfo{person}{W. Yin} {and} \bibinfo{person}{H.
  Sch{\"{u}}tze}.} \bibinfo{year}{2015}\natexlab{}.
\newblock \showarticletitle{Convolutional Neural Network for Paraphrase
  Identification}. In \bibinfo{booktitle}{\emph{NAACL '15}}.
  \bibinfo{pages}{901--911}.
\newblock


\bibitem[\protect\citeauthoryear{Yu, Qiu, Jiang, Huang, Song, Chu, and Chen}{Yu
  et~al\mbox{.}}{2018}]%
        {alime-tl}
\bibfield{author}{\bibinfo{person}{J. Yu}, \bibinfo{person}{M. Qiu},
  \bibinfo{person}{J. Jiang}, \bibinfo{person}{J. Huang}, \bibinfo{person}{S.
  Song}, \bibinfo{person}{W. Chu}, {and} \bibinfo{person}{H. Chen}.}
  \bibinfo{year}{2018}\natexlab{}.
\newblock \showarticletitle{Modelling Domain Relationships for Transfer
  Learning on Retrieval-based Question Answering Systems in E-commerce}. In
  \bibinfo{booktitle}{\emph{WSDM '18}}. \bibinfo{pages}{682--690}.
\newblock


\bibitem[\protect\citeauthoryear{Zhang, Wei, and Zhou}{Zhang
  et~al\mbox{.}}{2019}]%
        {DBLP:journals/corr/abs-1905-06566}
\bibfield{author}{\bibinfo{person}{X. Zhang}, \bibinfo{person}{F. Wei}, {and}
  \bibinfo{person}{M. Zhou}.} \bibinfo{year}{2019}\natexlab{}.
\newblock \bibinfo{title}{{HIBERT:} Document Level Pre-training of Hierarchical
  Bidirectional Transformers for Document Summarization}.
\newblock
\newblock
\showeprint{1905.06566}


\end{thebibliography}
